\documentclass[aps,prd,twocolumn,showpacs,preprintnumbers,amsmath,amssymb]{revtex4}
\usepackage{graphicx}
\usepackage{dcolumn}
\usepackage{bm}
\usepackage{amssymb,amsmath}
\usepackage{latexsym}

\allowdisplaybreaks

\begin{document}

\title{Interaction of plane gravitational waves with a Fabry-Perot cavity\\
in the local Lorentz frame}
\author{Sergey P. Tarabrin}
\affiliation{Faculty of Physics, Moscow State University, Moscow,
119992, Russia} \email{tarabrin@phys.msu.ru}
\date{\today}

\begin{abstract}
    We analyze the interaction of plane '+'-polarized gravitational
    waves with a Fabry-Perot cavity in the local Lorentz frame
    of the cavity input mirror outside of the range of long-wave
    approximation with the force of radiation
    pressure taken into account. The obtained detector response signal is represented as
    a sum of two parts: (i) the phase shift due to displacement of the movable mirror under the
    influence of gravitational wave and the force of light pressure, and (ii)
    the phase shift due to direct interaction of gravitational wave with light wave
    inside the cavity. We obtain formula for the movable mirror law of motion paying
    close attention to the phenomena of optical rigidity, radiative friction
    and direct coupling of gravitational wave to light wave. Some
    issues concerning the detection of high-frequency gravitational
    waves and the role of optical rigidity in it are discussed.
    We also examine in detail special cases of optical resonance and small detuning
    from it and compare our results with the known ones.
\end{abstract}
\preprint{LIGO-P060055} 

\maketitle

\section{Introduction}\label{intro}
The search for gravitational waves (GWs) is now conducted with large
Earth-based Michelson-type laser interferometers where test masses
are separated by a distance of several kilometers
\cite{weiss,abramovici,detection}. Variations in the proper
distances between these test masses are measured with the light
wave. Traditionally interferometric GW detectors are analyzed in the
transverse-traceless (TT) gauge, where coordinate lines are chosen
to coincide with the geodesics of the freely falling test masses (if
initial velocities of test masses are equal to zero)
\cite{basics_of_gw_theory,thorne_clas_phys}. This means that the
coordinate frame is rigidly bounded with the \textit{inertial} test
masses. For a '+'-polarized plane gravitational wave metric in the
TT gauge has the usual linearized form \cite{wheel_mis_thorne}:
\begin{align}
    ds^2=&g_{\alpha\beta}(x)dx^\alpha dx^\beta=
    \left[\eta_{\alpha\beta}+h_{\alpha\beta}(x)\right]dx^\alpha
    dx^\beta\nonumber\\
    =&-c^2dt^2+(1+h)dx^2+(1-h)dy^2+dz^2.
    \label{TT_metric}
\end{align}
Here $h=h(t-z/c)$, $|h|\ll 1$ is the GW wavefunction,
$\eta_{\alpha\beta}=\textrm{diag}(-1,1,1,1)$ is the metric tensor of
flat space-time, and $h_{\alpha\beta}$ are small deviations from it.
Greek indices run over $0,1,2,3$ or $ct,x,y,z$; Latin indices run
over $1,2,3$ or $x,y,z$.

Geodesic equation in metric (\ref{TT_metric}) dictates that the test
mass coordinates are not changed by gravitational wave
\cite{thorne_clas_phys}, i.e. GW does not interact with the test
masses in the TT gauge. Nevertheless, it noticeably changes the
phase of electromagnetic wave (EMW for short) which propagates in
the region of space-time where gravitational wave is present. The
elementary solution of eikonal equation for a single arm of laser
interferometer \cite{thorne_clas_phys} states that the phase shift
is proportional to GW signal: $\delta\Psi(t)=k_0Lh(t)$, where $k_0$
is the wavenumber of electromagnetic wave and $L$ is the
interferometer arm length. In other words one can reduce the
analysis of a '+'-polarized GW interaction with detector to the
problem of EMW propagation in the medium with effective refraction
index $n^2(t)=1+h(t)$. This result is only valid in long-wave
approximation $L\ll\lambda_{\textrm{gw}}$, where
$\lambda_{\textrm{gw}}$ is the length of gravitational wave. In
general case the formula for phase shift is more complex but it is
still the solution of the eikonal equation. The significant
advantage of the TT gauge is the ease of going from long-wave
approximation to the general case \cite{thorne_clas_phys}.

However, the crucial requirement of the test masses inertiality
cannot be met in the real Earth-based experiments since there exists
a great amount of non-gravitational forces acting on the test
masses, such as seismic and thermal fluctuating forces, light
pressure resulting in ponderomotive rigidity
\cite{rig_brag_1,rig_brag_2,rigidity_1,rigidity_2,rakh_phd_thesis,rigidity_3,rigidity_4,
rigidity_5,rigidity_6,rigidity_7,rigidity_8} and radiative friction
\cite{rad_frict}. The latter two phenomena appear if Fabry-Perot
(FP) cavities are utilized in the interferometer as in LIGO and
VIRGO detectors \cite{detection}. Therefore, the reference frame of
the TT gauge can hardly be constructed for the Earth-based
detectors. Correspondingly, the results obtained within the bounds
of the TT gauge are only applicable when condition of the test
masses inertiality is met (the case of space GW antenna LISA
\cite{LISA_1,LISA_2} in which drag-free technique will be used).
Otherwise, additional analysis should be performed in order to
validate the TT-based consideration of non-geodesic motion. Clearly,
this will significantly complicate the whole problem.

In order to get rid of inertiality requirement when analyzing the
interaction of GW with a Fabry-Perot cavity and take
non-gravitational forces into account we should work in the local
Lorentz frame (LL gauge) of the cavity input mirror. Coordinate
transformation from the TT gauge to the LL gauge can be found in
Ref. \cite{thorne_clas_phys}:
\begin{align}
    ds^2=g_{\alpha\beta}(x)dx^\alpha dx^\beta
    =&-c^2dt^2+dx^2+dy^2+dz^2\nonumber\\
    &+\,\frac{1}{2c^2}\,\left(x^2-y^2\right)\ddot{h}\,
    (c\,dt-dz)^2,
    \label{metric_tensor}
\end{align}
where old notations for the coordinates are used since we are not
interested in the coordinates of the TT gauge anymore.

A simple analysis performed in Ref. \cite{thorne_clas_phys} shows
that in the LL gauge gravitational wave mainly interacts with the
test masses, resulting in the phase shift $\delta\Psi(t)=2k_0\delta
X(t)=k_0Lh(t)$, while direct interaction with the light wave leads
only to a small correction in long-wave approximation. Here $\delta
X(t)=(1/2)Lh(t)$ is the displacement of the test mass due to its
interaction with gravitational wave. In other words one can reduce
the analysis of a '+'-polarized GW interaction with detector to the
problem of EMW propagation in the volume without effective medium
(in long-wave approximation only!) but with movable boundaries.

However, the LL gauge has one great disadvantage
--- the complexity of going from long-wave approximation to the
general case. In this paper we are mostly interested in the latter
because still little is known about the particular features of
non-inertial test masses motion under the influence of gravitational
wave when condition $L\ll\lambda_{\textrm{gw}}$ is violated. In this
sense we are mostly interested in the phenomenon of optical rigidity
(or optical spring) and its role in detection of high frequency GW
signals. It is most likely that the optical strings will be
implemented in the future generations of the Earth-based detectors
(such as Advanced LIGO), allowing significant amplification of the
response to GW signals. General formulas for optical rigidity, test
masses motion and detector response signal outside of the range of
long-wave approximation were obtained in Ref.
\cite{rakh_phd_thesis}. Starting from the ideas of Ref.
\cite{local_observ} we propose an independent derivation of general
formulas and obtain a good agreement with results of Ref.
\cite{rakh_phd_thesis} on the one hand, and of
\cite{rigidity_3,rad_frict} on the other. Our calculations allow us
to take into account the relativistic correction to the force of
radiation pressure, known as radiative friction \cite{rad_frict},
which was not obtained in Ref. \cite{rakh_phd_thesis}. Using modern
notations we further analyze the GW- and radiation pressure-driven
motion of test masses, paying close attention to general
relativistic aspects of the problem. We also study in detail the
differences between the general case and its long-wave
approximation. It should be noted that we deal only with the optimal
GW polarization and detector orientation ('+'-polarized GW with one
of its principal axes parallel to the axis of FP cavity) since it is
enough for consideration of the radiation pressure-induced dynamics.

Motivation for our work is the following. Modern GW detectors,
utilizing Fabry-Perot cavities, with arm lengths of several
kilometers (4 km for LIGO and 3 km for VIRGO) are best suited for
detection of GW signals with frequencies that lie in the region of
$10^2$ to $10^3$ Hz. For them $L/\lambda_{\textrm{gw}}\sim
10^{-3}\div 10^{-2}\ll 1$ and long-wave approximation holds with
high accuracy. Today the most efforts are made to detect these
low-frequency signals which are emitted, for example, by binary
pulsars, rotating neutron stars, supernova explosions etc. However,
modern astrophysical and cosmological theories predict the existence
of sources of high-frequency ($f\geq 10^4$ Hz) GW signals, such as
relic gravitational wave background
\cite{relic_background_1,relic_background_2,relic_background_3}.
Some models, such as inflation, predict the GW spectrum of the
background with typical values of amplitude which are far beyond the
sensitivity of even advanced interferometers. However, the spectrum
of gravitational waves predicted by string cosmology
\cite{string_cosmology_1,string_cosmology_2,string_cosmology_3} has
rather big typical GW amplitude which rises with increasing
frequency, thus allowing the potential possibility of detection of
GWs with certain frequencies above $10^3$ Hz on the Earth-based
detectors. There is a strong evidence that modern and future
detectors are (or will be) capable of detecting signals at
frequencies about $10^4$ Hz. For example, in Refs.
\cite{high_frequency_1,high_frequency_2} (see also Refs.
\cite{high_frequency_3,high_frequency_4}) it was proposed that 4 km
LIGO detectors could detect 37.5 kHz signals, corresponding to
cavity free spectral range (FSR), for which
$L/\lambda_{\textrm{gw}}=1/2$, thus operating outside of long-wave
regime. And though in this case the response signal vanishes for
optimal polarization of gravitational wave \cite{angular_resp}, GW
signals can still be detected with enhanced sensitivity at multiples
of the cavity FSR for other orientations. However, the role of
radiation pressure effects in the detection of high-frequency GWs
when condition $L/\lambda_{\textrm{gw}}\ll1$ is violated (especially
in the vicinity of FSR and its multiples) has not been recognized up
to date. We propose the study of these effects, since the searches
for GW background are planned for the Advanced LIGO stage where
optical springs will be utilized.

This paper is organized as follows. In Sec. \ref{wave_eq} we write
down the wave equation for electromagnetic wave propagating in the
space-time of plane gravitational wave and obtain its solution  for
an arbitrary $h(t)$. In Sec. \ref{monochromatic} we examine the
special case of monochromatic $h(t)$. In Sec. \ref{fabry-perot} we
apply these results to consideration of the boundary problem for EMW
in a Fabry-Perot cavity with the movable mirror and obtain its
response signal to GWs. By neglecting the resonant features of FP
cavity in Sec. \ref{single_refl} we find excellent agreement between
our LL-based calculations and TT-based calculations performed in
Ref. \cite{thorne_clas_phys} for the case of single reflection. In
Sec. \ref{eq_of_mot} we write down full equation of motion for the
mirror paying close attention to the phenomena of optical rigidity,
radiative friction and GR corrections to the force of radiation
pressure. In Sec. \ref{law_of_mot} we find the mirror law of motion
and substitute it into the response signal. The obtained formulas
basically agree with results of Ref. \cite{rakh_phd_thesis}. We also
discuss the possibility of parametric excitation of additional
optical modes under the influence of GW signals with frequencies
near cavity FSR and its multiples. In Sec. \ref{special} we examine
special cases of optical resonance and small detuning from it in
general case and long-wave approximation. Excellent agreement with
the results obtained in Refs. \cite{rigidity_3,rad_frict} is found.
In the end we estimate the deposit of optical rigidity into the
response signal in the vicinity of FSR.

\section{Wave equation and its solution}\label{wave_eq}
We start from the second pair of Maxwell's equations without sources
in curved space-time \cite{field_theory}:
\begin{equation*}
    \frac{1}{\sqrt{-g}}\frac{\partial}{\partial x^\beta}
    \left(\sqrt{-g}F^{\alpha\beta}\right)=0.
\end{equation*}
Here $F_{\mu\nu}=\partial_\mu A_\nu-\partial_\nu A_\mu$,
$A^{\mu}=(A^0,A^1,A^2,A^3)$ is the 4-potential of electromagnetic
field, and $g=\det(g_{\mu\nu})\equiv -1$. Let us impose the Coulomb
gauge $A^0=0,\
\partial_1A^1+\partial_2A^2+\partial_3A^3=0$. Then
\begin{equation*}
    g^{\alpha\beta}\partial_\alpha\partial_\beta A^\mu=0.
\end{equation*}
In general $A^\mu=A^\mu(x,y,z,t)$. Let us restrict ourselves to
consideration of gravitational waves with frequencies
$f_{\textrm{gw}}\sim 10^2\div 10^5$ Hz, corresponding to wavelengths
$\lambda_{\textrm{gw}}\sim 10^3\div 10^6$ m. This is far larger than
the optical wavelength $\lambda_0\approx 1$ mkm and the radius of
the laser beam $r\approx 10$ cm. For simplicity we assume that
$A^\mu$ describes a linearly $z$-polarized plane light wave
traveling along the $x$-axis. Therefore, under all listed conditions
we have $A^\mu=(0,0,0,A)$ with $A=A(x,t)$ and the following scalar
wave equation:
\begin{equation}
    \frac{\partial^2A}{\partial x^2}-\frac{1}{c^2}\frac{\partial^2A}{\partial t^2}=
    \frac{1}{2}\,\frac{x^2}{c^2}\,\ddot{h}(t)\,\frac{1}{c^2}\frac{\partial^2A}{\partial t^2}.
    \label{wave_equation}
\end{equation}
The right side of this equation describes the direct interaction of
gravitational wave with light wave \cite{thorne_clas_phys}. In Ref.
\cite{local_observ} this effect was called the distributed
gravitational redshift.

It is convenient to solve Eq. (\ref{wave_equation}) using the method
of successive approximations since $h\ll 1$. We shall keep only the
zeroth and the first order in $h$ terms:
$A(x,t)=A^{(0)}(x,t)+A^{(1)}(x,t), \ |A^{(1)}|\sim|hA^{(0)}|\ll
|A^{(0)}|$. The solution of the zeroth order is the sum of plane
monochromatic waves traveling in positive and negative directions of
the $x$-axis. We denote ``positive'' wave with index '+' and
``negative'' wave with index '--':
\begin{align}
    &A^{(0)}(x,t)=A_+^{(0)}(x,t)+A_-^{(0)}(x,t),\nonumber\\
    &A_\pm^{(0)}(x,t)=A_{\pm0}e^{-i(\omega_0t\mp k_0x)}+\textrm{c.c.},
    \label{zeroth_order_solution}
\end{align}
where $k_0=\omega_0/c$. Amplitude and frequency are derived from
some initial and boundary problems and we shall keep them undefined
until Sec. \ref{fabry-perot}. The first order equation is:
\begin{equation}
    \frac{\partial^2A^{(1)}}{\partial x^2}-
    \frac{1}{c^2}\frac{\partial^2A^{(1)}}{\partial t^2}=
    \frac{1}{2}\,\frac{x^2}{c^2}\,\ddot{h}(t)\,
    \frac{1}{c^2}\frac{\partial^2A^{(0)}}{\partial t^2}.
    \label{1st_order_equation}
\end{equation}
The general solution of this equation can be represented as the sum
of ``positive'' and ``negative'' waves:
\begin{equation}
    A^{(1)}(x,t)=A_+^{(1)}(x,t)+A_-^{(1)}(x,t).
    \label{sum}
\end{equation}
Clearly, they can be treated independently. Remind, that
$g_{00}(0,t)=-1$ and therefore, we must demand that
\begin{equation}
    A_+^{(1)}(0,t)=A_-^{(1)}(0,t)=0.
    \label{initial_conditions}
\end{equation}
Physically these initial conditions mean that both the light waves
$A_{\pm}(x,t)$ experience no redshift at $x=0$, i.e. the solution of
full Eq. (\ref{wave_equation}) at $x=0$ is
$A_\pm(0,t)=A^{(0)}_\pm(0,t)=A_{\pm0}e^{-i\omega_0t}+A_{\pm0}^*e^{i\omega_0t}$.
The solution of the Cauchy problem
(\ref{1st_order_equation}--\ref{initial_conditions}) is obtained in
Appendix \ref{1st_ord_sol}:
\begin{align}
    &A^{(1)}_\pm(x,t)=A_{\pm0}g_\pm(x,t)e^{-i(\omega_0t\mp k_0x)}
    +{\textrm{c.c.}},\label{1st_order_solution}\\
    &g_\pm(x,t)=\int_{-\infty}^{+\infty}g_\pm(x,\Omega+\omega_0)
    e^{-i\Omega t}\,\frac{d\Omega}{2\pi},
    \nonumber
\end{align}
where
\begin{align*}
    g_\pm(&x,\Omega)\equiv
    \frac{1}{2}\,h(\Omega-\omega_0)\\
    &\times\Biggl\{x^2\xi(\Omega)\mp ix\eta(\Omega)-\zeta(\Omega)
    \biggl[1-e^{\mp i\left(k_0-\frac{\Omega}{c}\right)x}\biggr]\Biggr\}.
\end{align*}
and
\begin{subequations}
\begin{align}
    &\xi(\Omega)\equiv\frac{\omega_0^2}{c^2}\,\frac{\Omega-\omega_0}{\Omega+\omega_0},\label{xi}\\
    &\eta(\Omega)\equiv4\,\frac{\omega_0}{c}\,\frac{\omega_0^2}{(\Omega+\omega_0)^2},\label{eta}\\
    &\zeta(\Omega)\equiv2\,\frac{\omega_0^2(\Omega^2+3\omega_0^2)}
    {(\Omega-\omega_0)(\Omega+\omega_0)^3}.\label{zeta}
\end{align}
\end{subequations}

\section{Case of monochromatic gravitational wave}\label{monochromatic}
\begin{figure}[h]
\begin{center}
\includegraphics[scale=0.60]{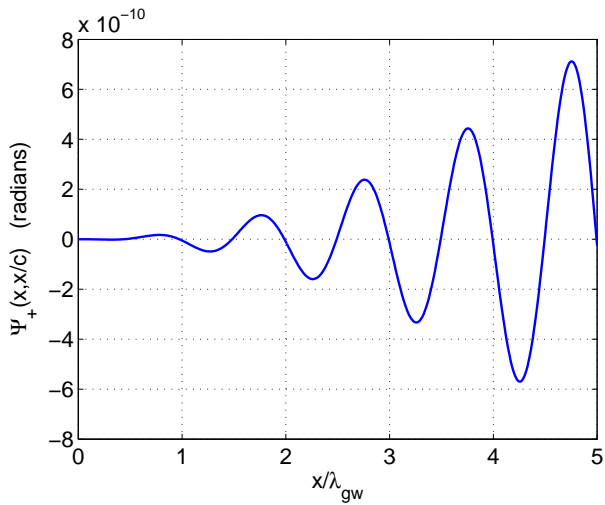}
\caption{Phase shift $\Psi_+(x,x/c)$ acquired by ``positive'' wave
from its direct interaction with monochromatic gravitational wave.
For visualization we assumed $\omega_0/2\pi=3\times10^{14}$ Hz,
$\omega_{\textrm{gw}}/2\pi=37.5$ kHz,
$h_0=10^{-22}$.}\label{pic_phase}
\end{center}
\end{figure}

In order to make clear the physical meaning of the obtained result
we shall examine the special case of monochromatic gravitational
wave $h(\Omega)=2\pi h_0\delta(\Omega-\omega_{\textrm{gw}})+2\pi
h_0\delta(\Omega+\omega_{\textrm{gw}})$, corresponding to
$h(t)=2h_0\cos\omega_{\textrm{gw}}t$ in time domain. We shall
restrict ourselves to consideration of the light wave traveling in
positive direction of the $x$-axis and use natural requirement
$\omega_{\textrm{gw}}/2\pi\sim 10^2\div 10^5\
\textrm{Hz}\ll\omega_0/2\pi\sim 10^{14}\ \textrm{Hz}$. Under these
conditions formulas (\ref{zeroth_order_solution}) and
(\ref{1st_order_solution}) lead to:
\begin{align*}
    A_+(x,t)&=\mathcal{A}_+(x,t)e^{i(k_0x-\omega_0t)}e^{i\Psi_+(x,t)}\\
    =A_{+0}\Biggl\{&1-ih_0\biggl[a(x)\sin\omega_{\textrm{gw}}t+b(x)\cos\omega_{\textrm{gw}}t\\
    &+\frac{\omega_0}{\omega_{\textrm{gw}}}\,
    \sin\omega_{\textrm{gw}}\left(t-\frac{x}{c}\right)\biggr]\Biggr\}
    e^{i(k_0x-\omega_0t)}+\textrm{c.c.},
\end{align*}
where
\begin{equation*}
    a(x)\equiv\frac{1}{2}\,\frac{x^2}{c^2}\,\omega_0\omega_{\textrm{gw}}-
    \frac{\omega_0}{\omega_{\textrm{gw}}},\qquad
    b(x)\equiv\frac{x}{c}\,\omega_0.
\end{equation*}
We can now easily extract information about the amplitude and the
phase of light wave. Up to the second order of $h$ amplitude remains
unchanged $\mathcal{A}_+(x,t)=A_{+0}$ while phase $\Psi_+(x,t)$ is
influenced by GW in the first order:
\begin{align*}
    \Psi_+(x,t)=
    &-h_0\Biggl[\biggl(\frac{x^2}{c^2}\,\omega_0\omega_{\textrm{gw}}-
    \frac{\omega_0}{\omega_{\textrm{gw}}}\biggr)\sin\omega_{\textrm{gw}}t\\
    &+\frac{x}{c}\,\omega_0\cos\omega_{\textrm{gw}}t+\frac{\omega_0}{\omega_{\textrm{gw}}}\,
    \sin\omega_{\textrm{gw}}\left(t-\frac{x}{c}\right)\Biggr].
\end{align*}
We must put $t=x/c$ if we are interested in the phase of light wave
at point $x$. This is enough since taking into account the influence
of gravitational wave on time will lead to the term of the second
order in $h$.
\begin{align*}
    \Psi_+(x,x/c)=-h_0\Biggl[\biggl(\frac{x^2}{c^2}\,\omega_0\omega_{\textrm{gw}}-
    &\frac{\omega_0}{\omega_{\textrm{gw}}}\biggr)
    \sin\omega_{\textrm{gw}}\,\frac{x}{c}\\
    &+\frac{x}{c}\,\omega_0\cos\omega_{\textrm{gw}}\,\frac{x}{c}\Biggr].
\end{align*}
We plotted the graph of this function in Fig. (\ref{pic_phase}).

The long-wave approximation $x/\lambda_{\textrm{gw}}\ll 1$ of this
formula is
\begin{align}
    \Psi_+(x)
    &\approx -\,\frac{2(2\pi)^2}{3}\,k_0xh_0\,
    \frac{x^2}{\lambda_{\textrm{gw}}^2}
    +\,o\left(k_0xh_0\,\frac{x^2}{\lambda_{\textrm{gw}}^2}\right)\nonumber\\
    &=O\left(k_0xh_0\,\frac{x^2}{\lambda_{\textrm{gw}}^2}\right).
    \label{phase_shift_long-wave}
\end{align}
This result coincides with the one of Ref. \cite{thorne_clas_phys}:
the direct interaction of gravitational wave and light wave leads
only to a small correction of the order of
$(x/\lambda_{\textrm{gw}})^2\,h$ in the LL gauge. Note that formula
(\ref{phase_shift_long-wave}) can be qualitatively interpreted as
the propagation of light wave in effective medium with refraction
index $n^2(x,t)=1+\delta n^2(x,t)$, where $\delta n(x,t)\propto
(x/c)^2\,\ddot{h}(t)$, $x\ll\lambda_{\textrm{gw}}$.

\section{Light wave in a Fabry-Perot cavity under the influence
of plane gravitational wave}\label{fabry-perot} 
\begin{figure}[h]
\begin{center}
\includegraphics[scale=0.55]{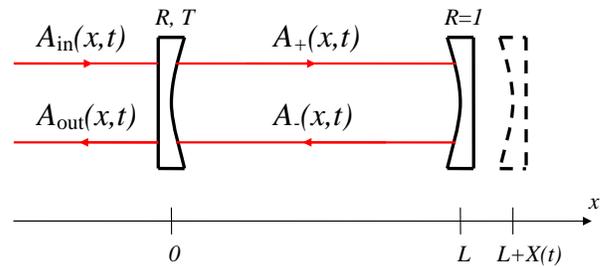}
\caption{Fabry-Perot cavity with the movable mirror. Input wave is
denoted by index 'in', output wave is 'out'. Optical field inside
the cavity is a sum of two waves running in two opposite directions
of the $x$-axis: ``positive'' wave is denoted by index '+' and
``negative'' is '--'. Isolated cavity without any gravitational or
optical radiation present has a fixed length $L$. Coordinate of the
movable mirror is $L+X(t),\ X\ll L$.} \label{pic_Fabry-Perot_cavity}
\end{center}
\end{figure}

Let us consider now
the light wave traveling inside the Fabry-Perot cavity placed in the
space-time of a plane '+'-polarized gravitational wave. The cavity
is formed by two mirrors (see Fig. \ref{pic_Fabry-Perot_cavity}).
The first one (left) is called the input mirror; it is immovable in
its own reference frame (remind that we work in the local Lorentz
frame of this mirror) and partially transmits optical radiation with
amplitude coefficient $T\ll 1$. The second mirror --- the end-mirror
--- is movable and absolutely reflective. For simplicity we assume
that both mirrors are ideal, meaning that they do not absorb or
dissipate optical energy. We put distance between the mirrors in the
absence of gravitational wave and optical radiation to be equal to
$L$.

Let the plane linearly polarized light wave $A_{\textrm{in}}(x,t)$
be incident on the input mirror. We shall write it down as a sum of
``strong'' wave with amplitude $A_{\textrm{in}0}$ and frequency
$\omega_0$ and ``weak'' addition $a_{\textrm{in}}(x,t)$ which
corresponds to the noises of light field:
\begin{align*}
        A_{\textrm{in}}(x,t)&=A_{\textrm{in}0}e^{-i(\omega_0t-k_0x)}+
        a_{\textrm{in}}(x,t)e^{-i(\omega_0t-k_0x)}+\textrm{c.c.},\\
        a_{\textrm{in}}(x,t)&=\int_{-\infty}^{+\infty}a_{\textrm{in}}(\Omega+\omega_0)
        e^{-i\Omega\left(t-\frac{x}{c}\right)}\,\frac{d\Omega}{2\pi}.
\end{align*}
For simplicity in this paper we consider classical EM fields (and
corresponding classical noises), since we are focused on dynamics
only. The generalization for quantized EM fields is straightforward:
classical field amplitudes should be replaced with corresponding
annihilation operators and complex conjugates --- with Hermitian
conjugates.

We represent the optical field inside the cavity as a sum of two
light waves: the one $A_+(x,t)$ running in the positive direction of
the $x$-axis and another one $A_-(x,t)$ running in the opposite
direction. We write them down as sums of three parts. The first part
is the ``strong'' wave with amplitude $A_{\pm0}$. The second part is
the explicit formula (\ref{1st_order_solution}) for direct coupling
'GW+EMW'. The third part is the ``weak'' addition $a_{\pm}(x,t)$
--- the unknown function of its arguments representing both optical
noises and phase shift acquired while circulating inside the cavity.
\begin{subequations}
\begin{align}
    A_\pm(x,t)=&A_{\pm0}e^{-i(\omega_0t\mp k_0x)}+A_{\pm0}g_\pm(x,t)e^{-i(\omega_0t\mp k_0x)}
    \nonumber\\
    &+a_\pm(x,t)e^{-i(\omega_0t\mp k_0x)}+{\textrm{c.c.}},\label{A_inside}\\
    a_\pm(x,t)=&\int_{-\infty}^{+\infty}a_\pm(\Omega+\omega_0)
    e^{-i\Omega\left(t\mp\frac{x}{c}\right)}\,\frac{d\Omega}{2\pi}.
    \label{a+-_forier_int}
\end{align}
\end{subequations}
We assume interaction between a gravitational wave $h$ and ``weak''
field $a_{\pm}$ to be a negligible effect since (i) it is most
likely that coupling of GWs to optical noises does not contain any
useful information and (ii) coupling of GWs to phase shift
(variation of optical field) acquired due to GW itself leads to the
negligible $h^2$ terms.

Output wave $A_{\textrm{out}}(x,t)$ includes the wave reflected from
the input mirror and the wave flowing out of the cavity. The latter
one carries information about gravitational wave (signal wave) along
with optical noises. Again we represent the wave as a sum of
``strong'' and ``weak'' components:
\begin{align*}
    A_{\textrm{out}}(x,t)=&A_{\textrm{out}0}e^{-i(\omega_0t+k_0x)}+
    a_{\textrm{out}}(x,t)e^{-i(\omega_0t+k_0x)}\\
    &+\textrm{c.c.},\\
    a_{\textrm{out}}(x,t)&=\int_{-\infty}^{+\infty}a_{\textrm{out}}(\Omega+\omega_0)
    e^{-i\Omega\left(t+\frac{x}{c}\right)}\,\frac{d\Omega}{2\pi}.
\end{align*}

Our problem is to find connection between $A_{\pm0},\
A_{\textrm{out}0},\ a_\pm(x,t),\ a_{\textrm{out}}(x,t)$ and
$A_{\textrm{in}0},\ a_{\textrm{in}}(x,t),\ g_\pm(x,t)$. This
connection is derived from the boundary conditions. The first
boundary condition is the requirement of vector potential continuity
along the input mirror:
\begin{align}
    A_{\textrm{out}}(0,\,t)&=RA_{\textrm{in}}(0,\,t)+TA_-(0,\,t),\label{boundary_1}\\
    A_+(0,\,t)&=TA_{\textrm{in}}(0,\,t)-RA_-(0,\,t).
    \label{boundary_2}
\end{align}
The second condition is the requirement of electric field to be
equal to zero on the surface of end-mirror in its local reference
frame \cite{mirror_radiation}: $E'_z(0,\,t-\tau)=0$ (we neglect
${\dot{X}}^2/c^2$-terms), where notation $\tau=L/c$ is introduced.
Using well-known formulas for Lorentz transformations of electric
and magnetic fields from movable frame of reference to the
laboratory frame \cite{field_theory} we rewrite this condition in
the following form:
\begin{equation}
    E_z(L+X,\,t-\tau)+(\dot{X}/c)H_y(L+X,\,t-\tau)=0,
    \label{electric_field_movable}
\end{equation}
or, substituting here definitions $E_z=-(1/c)(\partial A/\partial
t)$ and $H_y=-\partial A/\partial x$, we come to the following
equation:
\begin{equation*}
    -\frac{1}{c}\left[\frac{\partial A(L+X,\,t-\tau)}{\partial t}+
    \dot{X}\frac{\partial A(L+X,\,t-\tau)}{\partial x}\right]=0,
\end{equation*}
which means that directional derivative of vector-potential equals
to zero along the mirror trajectory. Thus it is sufficient to demand
$A(L+X,\,t-\tau)=0$ or
\begin{equation}
    A_+(L+X,\,t-\tau)+A_-(L+X,\,t-\tau)=0.
    \label{boundary_3}
\end{equation}
Here $X=X(t-\tau)$. We would like to stress that this boundary
condition automatically allows us to take into account the terms
proportional to $\dot{X}/c$ which lead to the appearance of the
force of radiative friction (see Sec. \ref{eq_of_mot}).

Strictly speaking, we cannot use Lorentz transformations to change
the reference frame in our case since space-time is curved. However,
one can verify that corrections to formulas of Lorentz
transformations due to gravitational wave are negligible.

We assume further that mirror's displacement $X(t)\sim Lh\ll L$.
Splitting the set of equations (\ref{boundary_1}, \ref{boundary_2},
\ref{boundary_3}) into the zeroth and the first order sets we obtain
their solutions. It should be noted here that we have not taken into
account the influence of GW on time in boundary conditions. For the
zeroth order this effect vanishes since dependence on time is only
included in the phase factor $e^{-i\omega_0t}$ which is a common
multiplier for both the left and the right sides of equations. For
the first order this effect is negligible since it will lead to
terms of the second order in $h$. Therefore, the zeroth order
solution is:
\begin{subequations}
\begin{align}
    &A_{+0}=\frac{T}{1-Re^{2i\omega_0\tau}}\,A_{\textrm{in}0},
    \label{A+_Ain}\\
    &A_{-0}=-\frac{Te^{2i\omega_0\tau}}{1-Re^{2i\omega_0\tau}}\,A_{\textrm{in}0},
    \label{A-_Ain}\\
    &A_{\textrm{out}0}=\frac{R-e^{2i\omega_0\tau}}{1-Re^{2i\omega_0\tau}}\,A_{\textrm{in}0}
    \nonumber,
\end{align}
\end{subequations}
and the first order solution in spectral domain is:
\begin{widetext}
\begin{subequations}
\begin{equation}
    a_+(\Omega+\omega_0)=
    a_{\textrm{in}}(\Omega+\omega_0)\,\frac{T}{1-Re^{2i(\Omega+\omega_0)\tau}}
    +\,\frac{Re^{2i\omega_0\tau}A_{+0}}{1-Re^{2i(\Omega+\omega_0)\tau}}\,i
    \Bigl[\delta\Psi_\textrm{mir}(\Omega)+\delta\Psi_{\textrm{gw+emw}}(\Omega)\Bigr],
    \label{a_plus}
\end{equation}
\begin{equation}
    a_-(\Omega+\omega_0)=-a_{\textrm{in}}(\Omega+\omega_0)
    \,\frac{Te^{2i(\Omega+\omega_0)\tau}}{1-Re^{2i(\Omega+\omega_0)\tau}}
    +\,\frac{A_{-0}}{1-Re^{2i(\Omega+\omega_0)\tau}}\,i
    \Bigl[\delta\Psi_\textrm{mir}(\Omega)+\delta\Psi_{\textrm{gw+emw}}(\Omega)\Bigr],
    \label{a_minus}
\end{equation}
\begin{equation}
    a_{\textrm{out}}(\Omega+\omega_0)=
    a_{\textrm{in}}(\Omega+\omega_0)\,\frac{R-e^{2i(\Omega+\omega_0)\tau}}
    {1-Re^{2i(\Omega+\omega_0)\tau}}
    +\,\frac{TA_{-0}}{1-Re^{2i(\Omega+\omega_0)\tau}}\,i
    \Bigl[\delta\Psi_\textrm{mir}(\Omega)+\delta\Psi_{\textrm{gw+emw}}(\Omega)\Bigr].
    \label{signal}
\end{equation}
\end{subequations}
\end{widetext}
Here we introduced the convenient notations for phase shifts after
single reflection:
\begin{align}
    \delta\Psi_\textrm{mir}(\Omega)=2k_0X(\Omega)e^{i\Omega\tau},
    \label{phase_shift_mir}
\end{align}
is the phase shift acquired by the light wave when reflecting from
the movable mirror. Note, that in time domain it comes with delay
$t-\tau$ since it takes light wave time $\tau$ to travel from the
end-mirror to the input mirror;
\begin{equation}
    \delta\Psi_{\textrm{gw+emw}}(\Omega)=
    i\Bigl[g_-(L,\Omega+\omega_0)-g_+(L,\Omega+\omega_0)\Bigr]e^{i\Omega\tau},
    \label{phase_shift_gw_emw}
\end{equation}
is the phase shift acquired by the light wave from its direct
interaction with gravitational wave. Below we calculate it
explicitly and show that in long-wave approximation this phase shift
vanishes.

\section{Phase shift of light wave in the case of single reflection}\label{single_refl}
Let us compare our results with the ones already described in
literature for the case of single reflection. To do this we ignore
all resonant features of our system. First, we shall rewrite our
results in spectral domain explicitly. For direct coupling of
gravitational wave to light wave (\ref{phase_shift_gw_emw}), taking
into account $\omega_{\textrm{gw}}\ll\omega_0$, we get:
\begin{equation}
    \delta\Psi_{\textrm{gw+emw}}(\Omega)\approx
    -k_0Lh(\Omega)\left(1-\frac{\sin\Omega\tau}{\Omega\tau}\right)e^{i\Omega\tau}.
    \label{direct_interaction}
\end{equation}
In time domain we consider the case of a monochromatic GW with
spectrum $h(\Omega)=2\pi h_0\delta(\Omega-\omega_{\textrm{gw}})+2\pi
h_0\delta(\Omega+\omega_{\textrm{gw}})$. After simple integration of
formula (\ref{direct_interaction}) we obtain:
\begin{align*}
    \delta\Psi_{\textrm{gw+emw}}(t)
    =&-2k_0Lh_0\cos\omega_{\textrm{gw}}(t-\tau)\nonumber\\
    &+h_0\,\frac{\omega_0}{\omega_{\textrm{gw}}}\,
    \biggl[\sin\omega_{\textrm{gw}}t-\sin\omega_{\textrm{gw}}(t-2\tau)\biggr].
\end{align*}
It is straightforward to verify that in long-wave approximation
$L/\lambda_{\textrm{gw}}\ll 1$ this formula reduces to
\begin{align*}
    \delta\Psi_{\textrm{gw+emw}}(t)\approx &-\,\frac{(2\pi)^2}{3}\,k_0Lh_0\,
    \left(\frac{L}{\lambda_{\textrm{gw}}}\right)^2\cos\omega_{\textrm{gw}}t\\
    &+o\left[k_0Lh_0\,\left(\frac{L}{\lambda_{\textrm{gw}}}\right)^2\right].
\end{align*}

Let us now turn to the full phase shift
$\delta\Psi_\textrm{LL}=\delta\Psi_\textrm{mir}+\delta\Psi_{\textrm{gw+emw}}$.
Until we find mirror law of motion we do not know how displacement
$X(\Omega)$ depends on gravitational wave $h(\Omega)$. However, for
single reflection (the case when we can ignore the force of light
pressure) this solution has been known long before:
$X(\Omega)=\frac{1}{2}Lh(\Omega)$. Therefore, the full signal is
\begin{equation}
    \delta\Psi_\textrm{LL}(\Omega)=
    k_0Lh(\Omega)\,\frac{\sin\Omega\tau}{\Omega\tau}\,e^{i\Omega\tau}.
    \label{full_signal}
\end{equation}
In time domain for monochromatic GW we get:
\begin{equation*}
    \delta\Psi_{\textrm{LL}}(t)=
    h_0\,\frac{\omega_0}{\omega_{\textrm{gw}}}\,
    \biggl[\sin\omega_{\textrm{gw}}t-\sin\omega_{\textrm{gw}}(t-2\tau)\biggr].
\end{equation*}
This result coincides with the one obtained in the framework of the
TT gauge in Ref. \cite{thorne_clas_phys}:
\begin{align*}
    \delta\Psi_{\textrm{TT}}&=-\frac{\omega_0}{2}\Biggl\{
    \int_0^{t-2\tau}h(t')\,dt'-\int_0^{t}h(t')\,dt'\Biggl\}\nonumber\\
    &=h_0\,\frac{\omega_0}{\omega_{\textrm{gw}}}\,
    \biggl[\sin\omega_{\textrm{gw}}t-\sin\omega_{\textrm{gw}}(t-2\tau)\biggr]
    =\delta\Psi_{\textrm{LL}}.
\end{align*}
This coincidence confirms the results of Ref. \cite{local_observ}:
formulas for the net phase shift obtained in the TT and the LL
gauges are equivalent if only in LL gauge one takes into account the
effects of localized and distributed gravitational redshifts (along
with displacement of mirror). Remind that distributed redshift is
the direct interaction of gravitational wave with light wave. Note
that we have not considered the effect of localized redshift simply
because our clocks are located at $x=0$ where $g_{00}(0,t)=-1$. For
generalization one may turn to Ref. \cite{local_observ}.

Again it is straightforward to verify that in long-wave
approximation the full signal reduces to
\begin{equation*}
    \delta\Psi_\textrm{LL}(t)\approx
    2k_0Lh_0\cos\omega_{\textrm{gw}}t+
    O\left(k_0Lh_0\,\frac{L}{\lambda_{\textrm{gw}}}\right),
\end{equation*}
thus leading to the rising of the term containing
$(L/\lambda_{\textrm{gw}})\,h_0$ in the full phase shift. This was
not mentioned in Ref. \cite{thorne_clas_phys} since the delay
$t-\tau$ of the reflected light wave was not taken into account in
$\delta\Psi_\textrm{mir}$ there:
$\delta\Psi_\textrm{mir}(t)=2k_0X(t-\tau)$ but not $2k_0X(t)$.

\section{Mirror equation of motion}\label{eq_of_mot}
In the absence of non-gravitational forces test mass moves along the
geodesic (with respect to the input mirror of the cavity where the
origin of coordinate frame is set) as dictated by geodesic equation
\cite{wheel_mis_thorne,field_theory}. It is straightforward to
verify that Christoffel coefficients are
$\Gamma^1_{00}=-x\ddot{h}/2c^2$, $\Gamma^1_{01}=\Gamma^1_{11}=0$
(for the motion along the $x$-axis) in metric (\ref{metric_tensor}),
so in non-relativistic limit we obtain the following equation of
motion for a free mass:
\begin{equation}
    \frac{d^2x}{dt^2}-\frac{1}{2}\,x\ddot{h}(t)=0.
    \label{geodesic_equation}
\end{equation}
We stress here that this equation is strict for any separation
between the test masses. Its solution is trivial if obtained with
the method of successive approximations. By assuming
$x(t)=X^{(0)}(t)+X^{(1)}(t)$ with $X^{(0)}=L$ we get
$X^{(1)}=X(t)=\frac{1}{2}\,Lh(t)$.

Generalization of Eq. (\ref{geodesic_equation}) for the case of
non-geodesic motion is the following:
\begin{equation*}
    \frac{d^2x}{dt^2}-\frac{1}{2}\,x\ddot{h}(t)=
    \frac{1}{m}\,F_x(x,\dot{x},t),
\end{equation*}
where $F$ is the force of non-gravitational nature, the force of
radiation pressure in our case, and $m$ is the test mass.

In general, the force of light pressure should be calculated in
curved space-time by means of General Relativity. However in
Appendix \ref{GR_corrections} we show that this calculation can be
performed in the framework of classical flat space-time
electrodynamics. We should explain this statement in detail. The
rule for calculation of the light pressure force in flat space-time
for any given electromagnetic wavefunction $A(x,t)$ is known
\cite{field_theory}. Obviously, there exists a general rule for
calculation of light pressure on curved space-time background. Our
statement is that this general ``curved space-time rule'' differs
from ``flat space-time rule'' only in negligible corrections.
Therefore, in our case we can apply the ``flat space time rule'' to
$A(x,t)$ which includes gravitational wavefunction $h$ (via direct
coupling of GW to EMW) without introducing any significant error.
Namely, in Appendix \ref{GR_corrections} we estimate the resulting
inaccuracy as $\delta
X^{(1)}_{\textrm{GR}}=F^{(1)}_{\textrm{GR}}/m\Omega^2\approx
(\mathcal{E}_{\textrm{FP}}/mc^2)Lh(t)\approx 10^{-17}Lh(t)$ (here
$\mathcal{E}_{\textrm{FP}}$ is the full energy of light wave inside
the cavity) compared to the first order solution
$X^{(1)}=\frac{1}{2}\,Lh(t)$ of Eq. (\ref{geodesic_equation}),
assuming that $\mathcal{E}_{\textrm{FP}}=20$ J and $m=40$ kg
(typical parameters for Advanced LIGO project).

In classical electrodynamics \cite{field_theory} spatial components
of the radiation pressure force are $F^i=\int T^{ik}\,dS_k$, where
$T^{ik}$ are the components of the energy-stress tensor of
electromagnetic field and $dS_k$ is the oriented area element. Since
the light beam with cross-section $S$ is parallel to the $x$-axis
and normal to the mirror surface then the $x$-component of the light
pressure force is $F_x=T^{11}S=T_{11}S$, where
$T_{11}=(E_z^2+H_y^2)/8\pi$, $E_z=-(1/c)(\partial A/\partial t)$ and
$H_y=-\partial A/\partial x$. Therefore, we obtain the following
first order equation of motion:
\begin{equation}
    m\left[\frac{d^2X}{dt^2}-\frac{1}{2}\,L\ddot{h}(t)\right]=
    F_x^{(1)}(X,\dot{X},t),\label{equation_of_motion}
\end{equation}
where
\begin{equation*}
    F_x^{(1)}(X,\dot{X},t)=
    \frac{S}{8\pi}\left[\frac{1}{c^2}\left(\frac{\partial A}{\partial t}\right)^2+
    \left(\frac{\partial A}{\partial x}\right)^2\right]^{(1)}_{x=L+X(t)}.
\end{equation*}
Index $(1)$ on the right side of equation means that only the first
order in $h$ and $X$ terms should be considered. Note that the
zeroth order force of light pressure is constant, therefore, we
should redefine the length $L$ in all formulas, assuming that in
real experiment constant force is compensated somehow. The first
order force is calculated explicitly in spectral domain in Appendix
\ref{force} (below we use notation $F_x$ instead of $F_x^{(1)}$):
\begin{equation*}
    F_x(\Omega)=
    F_{\textrm{fluct}}(\Omega)+F_{\textrm{pm}}(\Omega)+F_{\textrm{gw+emw}}(\Omega),
\end{equation*}
where
\begin{align*}
    F_{\textrm{fluct}}(\Omega)&\\
    =\frac{Sk_0}{\pi c}\Biggl[
    &\frac{T^2A^*_{\textrm{in}0}}{1-Re^{-2i\omega_0\tau}}\,
    \frac{(\omega_0+\Omega)e^{i\Omega\tau}}{1-Re^{2i(\omega_0+\Omega)\tau}}\,
    a_{\textrm{in}}(\omega_0+\Omega)\\
    &+\frac{T^2A_{\textrm{in}0}}{1-Re^{2i\omega_0\tau}}\,
    \frac{(\omega_0-\Omega)e^{i\Omega\tau}}{1-Re^{-2i(\omega_0-\Omega)\tau}}\,
    a^*_{\textrm{in}}(\omega_0-\Omega)\Biggr],
\end{align*}
is the fluctuation part of the force (see Appendix
\ref{force_fluct}), which leads to the quantum radiation pressure
noise (back-action) after the quantization of the light field;
\begin{align*}
    F_{\textrm{pm}}(\Omega)=&-\mathcal{K}(\Omega)X(\Omega)\\=
    &-K(\Omega)X(\Omega)+2i\Omega\,\Gamma(\Omega)X(\Omega),
\end{align*}
is the ponderomotive force, where (see Appendix
\ref{force_optmech_couple}):
\begin{equation}
    K(\Omega)=\frac{4k_0SW_{\textrm{FP}}Re^{2i\Omega\tau}\sin2\omega_0\tau}
    {1-2Re^{2i\Omega\tau}\cos2\omega_0\tau+R^2e^{4i\Omega\tau}},
    \label{optical_rigidity}
\end{equation}
is the coefficient of optical rigidity and
\begin{align}
    \Gamma(\Omega)=\frac{SW_{\textrm{FP}}}{c}\,\frac{1-R^2e^{4i\Omega\tau}}
    {1-2Re^{2i\Omega\tau}\cos2\omega_0\tau+R^2e^{4i\Omega\tau}},
    \label{radiation_friction}
\end{align}
is the coefficient of radiative friction. Here
$W_{\textrm{FP}}=k_0^2(A_{+0}A^*_{+0}+A_{-0}A^*_{-0})/2\pi$ is the
energy density of light wave circulating inside the cavity. Formula
(\ref{optical_rigidity}) was obtained in Refs.
\cite{rigidity_3,rad_frict,rakh_phd_thesis} and formula
(\ref{radiation_friction}) in Ref. \cite{rad_frict}, where
phenomenon of radiative friction was discussed in detail.

We split up $\mathcal{K}$ into $K$ and $2i\Omega\Gamma$ using the
following reasoning. The obtained mechanical impedance $Z(\Omega)$
of the mirror is
$Z(\Omega)=m\Omega^2+2i\Omega\,\Gamma(\Omega)-K(\Omega)$ which
obviously looks similar to the impedance of linear harmonic
oscillator. Therefore, $\mathcal{K}(\Omega)$ includes the term
$2i\Omega\,\Gamma(\Omega)$ describing friction, and the term
$K(\Omega)$ describing rigidity, where $2i\Omega\Gamma$ is of the
order of $(\Omega/\omega_0)K$. Radiative friction $\Gamma(\Omega)$
appears due to the boundary condition (\ref{electric_field_movable})
which takes into account the terms proportional to $\dot{X}/c$. This
relativistic correction was not considered in Ref.
\cite{rakh_phd_thesis}.

Both the terms have real and imaginary parts. The latter means that
both effects are introduced with delay. Note that the coefficient of
``true'' friction is only the real part of formula
(\ref{radiation_friction}). ``True'' friction is always masked with
the imaginary part of formula (\ref{optical_rigidity}) since
$|K|\gg|2\Omega\Gamma|$, except for the case of optical resonance
(see later) when optical rigidity vanishes. In non-resonant regime
there exist two possibilities (in long-wave approximation),
depending on the sign of $\sin2\omega_0\tau$: either (i)
$\mathfrak{R}\bigl[K(\Omega)\bigr]>0$ (positive rigidity) and
$\mathfrak{I}\bigl[K(\Omega)\bigr]<0$ (negative damping), or (ii)
$\mathfrak{R}\bigl[K(\Omega)\bigr]<0$ (negative rigidity) and
$\mathfrak{I}\bigl[K(\Omega)\bigr]>0$ (positive damping). These
inequalities were first obtained in Refs.
\cite{rig_brag_1,rig_brag_2}.

The concept of ponderomotive rigidity and the role it plays in
standard quantum limited (SQL) detectors has been widely discussed
in literature before (for long-wave approximation only). For the
first time it appeared in Refs. \cite{rig_brag_1,rig_brag_2} where
microwave resonators were considered. Optical ponderomotive rigidity
and its application to large-scale interferometric GW detectors were
studied in Refs.
\cite{rigidity_1,rigidity_2,rigidity_3,rakh_phd_thesis}. Papers
\cite{rigidity_4,rigidity_5,rigidity_6,rigidity_7,rigidity_8} dealt
with the optical springs in signal-recycling topologies of LIGO
detectors;
\begin{equation}
    F_{\textrm{gw+emw}}(\Omega)=
    \frac{1}{2}\,\mathcal{K}(\Omega)
    \left(1-\frac{\sin\Omega\tau}{\Omega\tau}\right)Lh(\Omega),
    \label{gw_emw_correction}
\end{equation}
is the 'GW+EMW' correction to the force of radiation pressure (see
Appendix \ref{force_gw_emw_correction}), arising when the cavity
length becomes comparable with gravitational wavelength. Note, that
it is proportional to $\delta\Psi_{\textrm{gw+emw}}(\Omega)$ much as
ponderomotive force is proportional to
$\delta\Psi_{\textrm{mir}}(\Omega)$. Obviously, in long-wave
approximation $L\ll\lambda_{\textrm{gw}}$ or $\Omega\tau\ll1$ this
'GW+EMW' correction vanishes (along with the corresponding phase
shift). Similar formula (without $\Gamma$ in $\mathcal{K}$) was
obtained in Ref. \cite{rakh_phd_thesis}.

\section{Mirror's law of motion and detector's response signal}\label{law_of_mot}
Now we solve Eq. (\ref{equation_of_motion}) in spectral domain:
\begin{align}
    X(\Omega)=
    \frac{1}{2}\,\frac{m\Omega^2Lh(\Omega)}{m\Omega^2-\mathcal{K}(\Omega)}-
    \frac{F_{\textrm{gw+emw}}(\Omega)+F_{\textrm{fluct}}(\Omega)}
    {m\Omega^2-\mathcal{K}(\Omega)}.
    \label{displacement_spectral}
\end{align}
In this paper we are mostly interested in dynamical part of the
problem, therefore, we shall ignore fluctuations (terms containing
$a_{\textrm{in}}$ and $F_{\textrm{fluct}}$) further. Substituting
(\ref{gw_emw_correction}) into (\ref{displacement_spectral}) we
obtain:
\begin{equation}
    X(\Omega)=\frac{1}{2}\,\frac{Lh(\Omega)}{m\Omega^2-\mathcal{K}(\Omega)}
    \Biggl[m\Omega^2-\mathcal{K}(\Omega)
    \biggl(1-\frac{\sin\Omega\tau}{\Omega\tau}\biggr)\Biggr].
    \label{law_of_motion}
\end{equation}
Strictly speaking, in order to keep all the $\dot{X}/c$-proportional
terms we should expand $g_{\pm}$ in formula
(\ref{phase_shift_gw_emw}) up to the first order of
$\Omega/\omega_0$ and then replace $(1-\textrm{sinc}\,\Omega\tau)$
with $(1-\Omega/\omega_0)(1-\textrm{sinc}\,\Omega\tau)$ in formulas
(\ref{gw_emw_correction}) and (\ref{law_of_motion}). However, for
$\Omega/2\pi\sim 10^2\div 10^5$ Hz we have $\Omega/\omega_0\sim
10^{-12}\div 10^{-9}$ and this correction can be omitted in both the
cases $K(\Omega)\neq0$ (term $K\times\Omega/\omega_0$ is negligible)
and $K(\Omega)=0$ (term
$2\Omega\Gamma\times\Omega/\omega_0\propto{\dot{X}}^2/c^2$ is
negligible). In addition, the term in square brackets in formula
(\ref{law_of_motion}) does not contain corrections of the order of
$(\mathcal{E}_{\textrm{FP}}/mc^2)m\Omega^2\approx 10^{-17}m\Omega^2$
since we have omitted them when calculating the radiation pressure
force.

Substituting formulas (\ref{direct_interaction}) and
(\ref{law_of_motion}) into (\ref{signal}) we finally obtain the
explicit formula for the optical field of the signal wave:
\begin{align}
    a_{\textrm{out}}(\Omega+\omega_0)=
    &-A_{\textrm{in}0}\frac{T^2e^{2i\omega_0\tau}}{\bigl(1-Re^{2i\omega_0\tau}\bigr)
    \bigl(1-Re^{2i(\Omega+\omega_0)\tau}\bigr)}\nonumber\\
    &\times\frac{m\Omega^2}{m\Omega^2-\mathcal{K}(\Omega)}
    \,ik_0Lh(\Omega)\,\frac{\sin\Omega\tau}{\Omega\tau}\,
    e^{i\Omega\tau}.
    \label{signal_wave}
\end{align}
Similarly to the previous formula one can replace
$\textrm{sinc}\,\Omega\tau$ with
$\bigl[1-(1-\Omega/\omega_0)\bigl(1-\textrm{sinc}\,\Omega\tau\bigr)\bigr]$
for strictness, but for the reasons discussed above this correction
is negligible.

Note that the obtained response signal is the product of three
terms: optical resonant multiplier, mechanical resonant multiplier
and the term (\ref{full_signal}) describing the single reflection
response. The latter is proportional to $\textrm{sinc}$-function and
is sometimes referred to as the ``electric'' component of the
detector full (single reflection) angular pattern
\cite{mag_comp_lw}. Recently it was demonstrated that the angular
pattern also contains the so-called ``magnetic'' component
\cite{mag_comp_lw} (due to non-optimal GW polarization and detector
orientation in general case) with its amplitude proportional to
$\Omega\tau$ in low frequencies but rising with increasing frequency
\cite{mag_comp_gen}. Therefore, it may be interesting to perform a
(straightforward) generalization of our results for the case of
generic GW polarization and direction of propagation. Probably, it
will add up to the multiplication of the known single reflection
angular pattern by the same optical and mechanical resonant
multipliers as in formula (\ref{signal_wave}) but, strictly
speaking, this problem requires a detailed analysis.

In many cases the phenomenon of radiative friction is negligible in
comparison with optical rigidity since $|2\Omega\Gamma|\ll K$. Then
one should replace $\mathcal{K}(\Omega)\rightarrow K(\Omega)$ in
formulas (\ref{law_of_motion}) and (\ref{signal_wave}).

\begin{figure}[h]
\begin{center}
\includegraphics[scale=0.55]{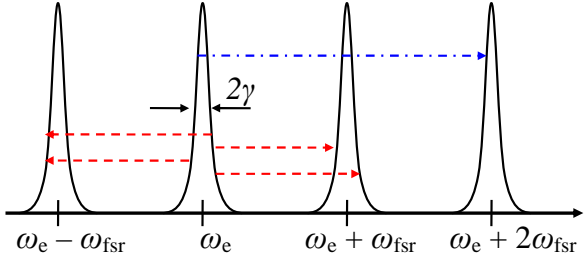}
\caption{Optical resonances of Fabry-Perot cavity:
$\omega_\textrm{e}$ is one of the eigenfrequencies,
$\omega_{\textrm{fsr}}=\pi c/L$ is the cavity free spectral range,
$\gamma=(1-R)/2\tau$ is cavity half-bandwidth. Red dashed arrows
denote some possible transitions between neighboring resonant curves
under the influence of GW with frequency equal or approximately
equal to FSR: $\omega_{\textrm{gw}}\approx \omega_{\textrm{fsr}}$.
Blue dash-dotted line denote one of the possible transitions between
non-neighboring curves if $\omega_{\textrm{gw}}\approx
2\omega_{\textrm{fsr}}$.} \label{pic_resonances}
\end{center}
\end{figure}

One of the most interesting effects which may arise when detecting
the high-frequency GW signal is parametric transitions between
optical modes of the cavity if $\Omega=\omega_{\textrm{gw}}\approx
n\omega_{\textrm{fsr}}$, where $n\in \mathbb{N}$ and
$\omega_{\textrm{fsr}}=\pi c/L$ is the cavity free spectral range
(see Fig. \ref{pic_resonances}). In some sense this effect looks
similar to the phenomenon of parametric oscillatory instability
(POI) described in Refs. \cite{param_instab_1,param_instab_2}. The
difference is in the following: POI appears due to the coupling of
optical modes to acoustic (mechanical) modes of the movable mirror
while the effect illustrated in Fig. \ref{pic_resonances} appears
due to the coupling of optical modes to gravitational wave itself.
This effect can be used for detection of high-frequency signals and
requires additional analysis.

\section{Special cases}\label{special}
It is useful to examine the obtained results in two special cases
--- optical resonance and small detuning from it --- and compare our
results with the ones already present in literature. We shall use
notations for the energy of light wave inside the cavity
$\mathcal{E}_{\textrm{FP}}=W_{\textrm{FP}}SL$, and cavity
half-bandwidth $\gamma=(1-R)/2\tau$. In these notations
$T^2=1-R^2\approx 4\gamma\tau$ and
$\mathcal{E}_{\textrm{FP}}=2P_{\textrm{in}}/\gamma$, where
$P_{\textrm{in}}$ is the input power.

\subsection{Optical resonance}
Let us assume that frequency of the pump wave $\omega_0$ coincides
with one of the eigenfrequencies of a Fabry-Perot cavity:
$\omega_0\tau=\pi n,\ n\in \mathbb{N}$. We shall denote all
quantities with index 'res' in this section. Then, simplifying
formulas (\ref{optical_rigidity}), (\ref{radiation_friction}),
(\ref{law_of_motion}) and (\ref{signal_wave}) we obtain
\begin{equation*}
    K^{\textrm{res}}(\Omega)=0,\qquad
    \Gamma^{\textrm{res}}(\Omega)=\frac{SW_{\textrm{FP}}}{c}\,\frac{1+Re^{2i\Omega\tau}}
    {1-Re^{2i\Omega\tau}},
\end{equation*}
thus in the case of resonance rigidity vanishes and radiative
friction is not masked by it;
\begin{align*}
    X^{\textrm{res}}(\Omega)=\frac{1}{2}\,&Lh(\Omega)
    \left[1-\frac{2i\Omega\,\Gamma^{\textrm{res}}(\Omega)}
    {m\Omega^2+2i\Omega\,\Gamma^{\textrm{res}}(\Omega)}
    \,\frac{\sin\Omega\tau}{\Omega\tau}\right],\\
    a^{\textrm{res}}_{\textrm{out}}(\Omega+\omega_0)
    =&-A_{\textrm{in}0}\,\frac{2}{1-Re^{2i\Omega\tau}}
    \frac{m\Omega^2}{m\Omega^2+2i\Omega\,\Gamma^{\textrm{res}}(\Omega)}\\
    &\times ik_0Lh(\Omega)\,\frac{\sin\Omega\tau}{\Omega\tau}\,
    e^{i\Omega\tau}.
\end{align*}
However, radiative friction is still a tiny effect even in the case
of optical resonance and is omitted usually:
\begin{align*}
    X^{\textrm{res}}(\Omega)&\approx\frac{1}{2}\,Lh(\Omega),\\
    a^{\textrm{res}}_{\textrm{out}}(\Omega+\omega_0)&\approx
    -A_{\textrm{in}0}\,\frac{2}{1-Re^{2i\Omega\tau}}\,
    ik_0Lh(\Omega)\,\frac{\sin\Omega\tau}{\Omega\tau}\,e^{i\Omega\tau}.
\end{align*}
Therefore, in the case of optical resonance light field inside the
cavity does not act dynamically on test mass if we neglect radiative
friction, i.e. test mass is inertial. Resonant features of the
cavity only appear in frequency-dependent amplification of the light
wave amplitude.

In long-wave approximation these formulas, denoted with 'lw', reduce
to the well-known ones:
\begin{equation*}
    K^{\textrm{res}}_{\textrm{lw}}(\Omega)=0,\qquad
    \Gamma^{\textrm{res}}_{\textrm{lw}}(\Omega)\approx
    \frac{\mathcal{E}_{\textrm{FP}}}{L^2}\,\frac{1+i\Omega\tau}{\gamma-i\Omega}.
\end{equation*}
The coefficient of ``true'' friction is the real part of
$\Gamma^{\textrm{res}}_{\textrm{lw}}$:
\begin{equation*}
    \mathfrak{R}\bigl[\Gamma^{\textrm{res}}_{\textrm{lw}}(\Omega)\bigr]
    =\frac{\mathcal{E}_{\textrm{FP}}}{L^2}\,\frac{\gamma}{\gamma^2+\Omega^2}=
    \frac{32P_{\textrm{in}}}{c^2}\,\frac{1}{T^4+(4\Omega\tau)^2}.
\end{equation*}
Last formula with $P_{\textrm{in}}$ coincides with the one in Ref.
\cite{rad_frict};
\begin{align*}
    X^{\textrm{res}}_{\textrm{lw}}&(\Omega)=\frac{1}{2}\,Lh(\Omega)\,
    \frac{m\Omega^2}{m\Omega^2+2i\Omega\,\Gamma^{\textrm{res}}_{\textrm{lw}}(\Omega)}
    \approx \frac{1}{2}\,Lh(\Omega),\\
    a^{\textrm{res}}_{{\textrm{out, lw}}}&(\Omega+\omega_0)\\
    &\approx -A_{\textrm{in}0}\frac{1}{(\gamma-i\Omega)\tau}
    \frac{m\Omega^2}{m\Omega^2+2i\Omega\,\Gamma^{\textrm{res}}_{\textrm{lw}}(\Omega)}\,
    ik_0Lh(\Omega)\\
    &\approx -A_{\textrm{in}0}\frac{1}{(\gamma-i\Omega)\tau}\,
    ik_0Lh(\Omega).
\end{align*}

\subsection{Small detuning from optical resonance}
\begin{figure}[h]
\begin{center}
\includegraphics[scale=0.60]{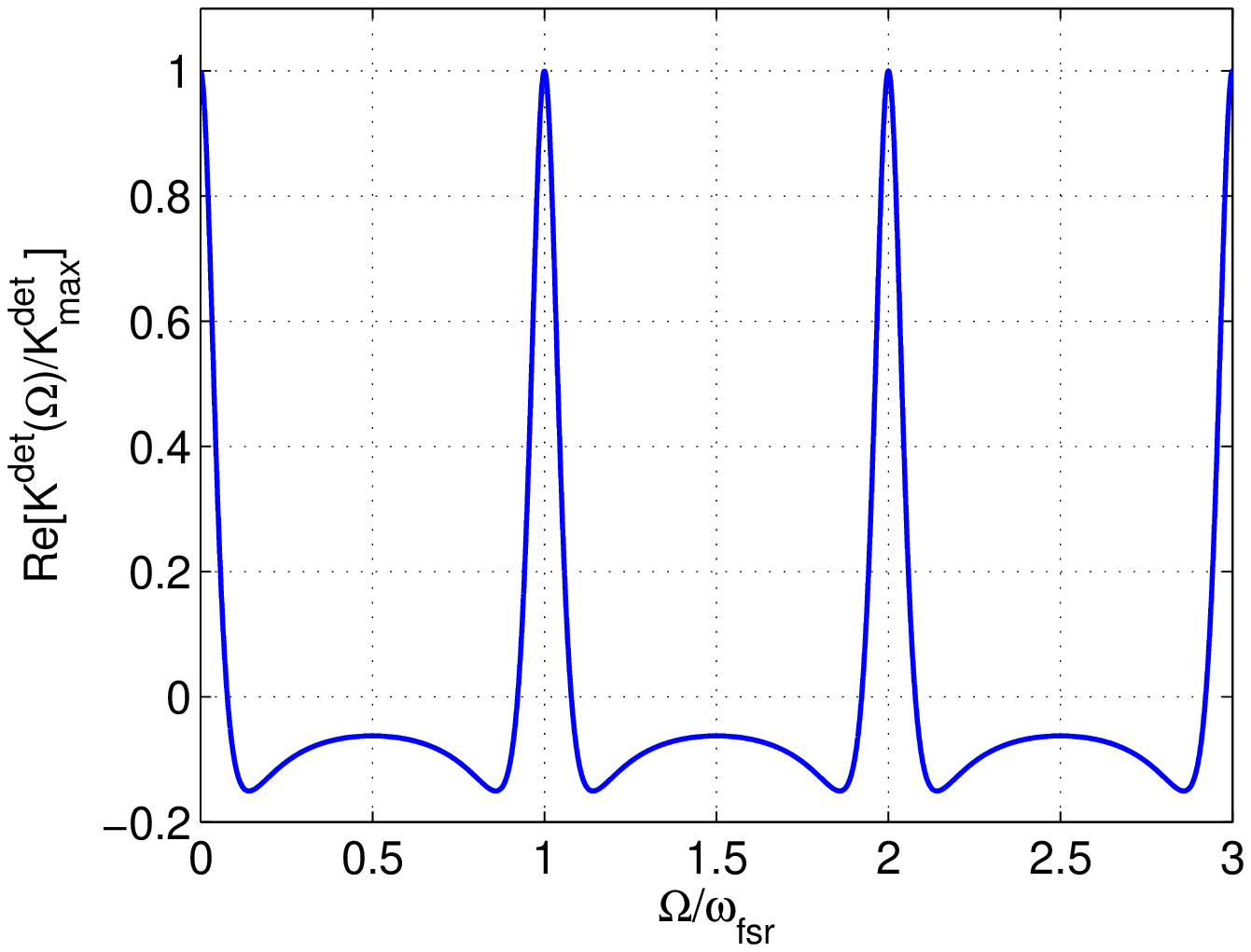}
\includegraphics[scale=0.60]{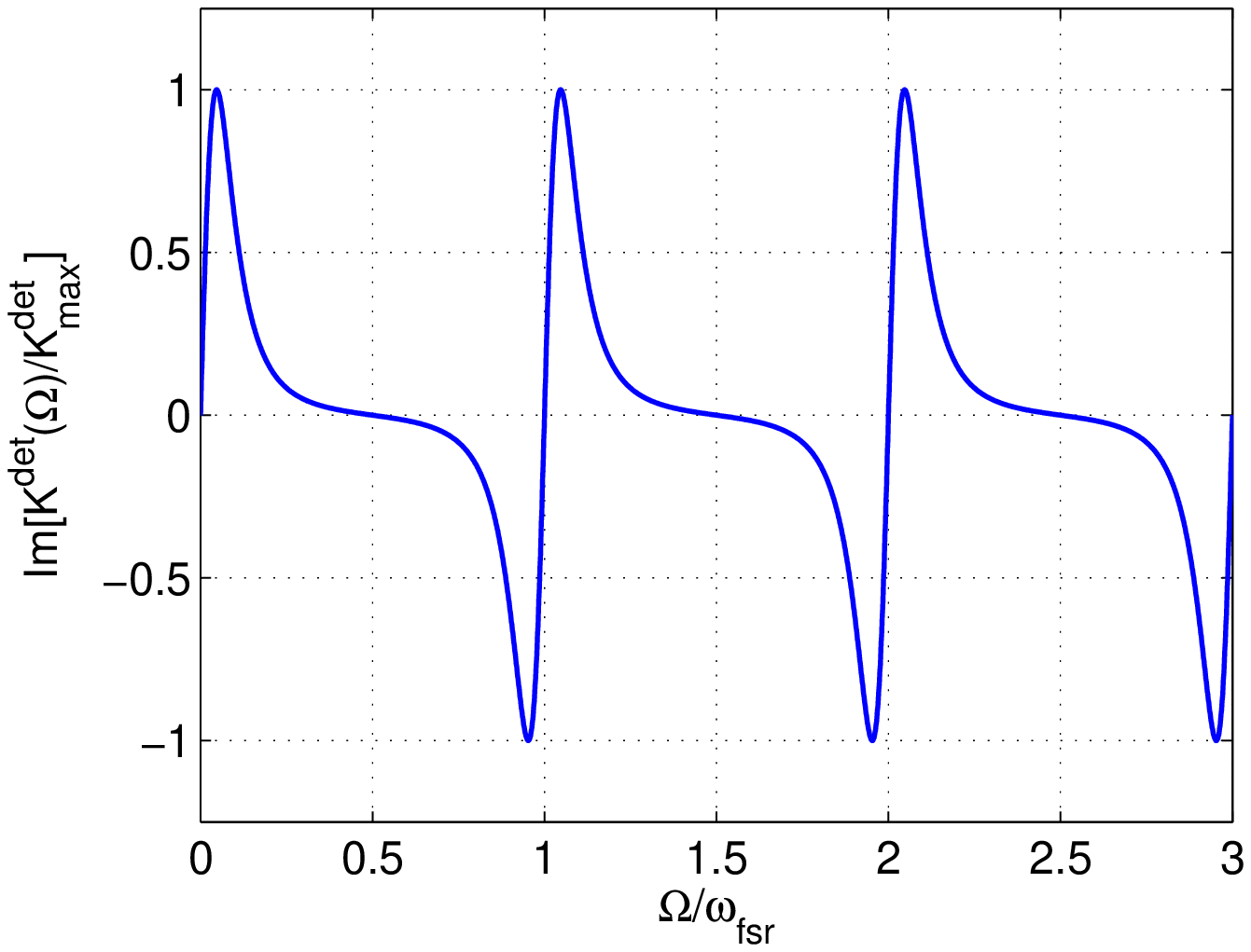}
\caption{``True'' rigidity (upper figure) and damping (lower figure)
normalized to their maximum values. For visualization we assumed
$R=0.6$, $\delta/2\pi=100$ Hz.}\label{pic_K}
\end{center}
\end{figure}

We shall now turn to the case when frequency of the pump wave is
slightly shifted from one of the eigenfrequencies:
\begin{equation*}
    \omega_0=\frac{\pi n}{\tau}+\delta,\qquad \delta\tau\ll 1.
\end{equation*}
Again, simplifying formulas (\ref{optical_rigidity}),
(\ref{law_of_motion}) and (\ref{signal_wave}) we obtain (all
quantities in this section are marked with 'det'):
\begin{equation}
    K^{\textrm{det}}(\Omega)=\frac{8k_0SW_{\textrm{FP}}Re^{2i\Omega\tau}\delta\tau}
    {(1-Re^{2i\Omega\tau})^2+4R(\delta\tau)^2e^{2i\Omega\tau}}.
    \label{detuning_rigidity}
\end{equation}
The graphs of $\mathfrak{R}\bigl[K^{\textrm{det}}(\Omega)\bigr]$
(``true'' rigidity) and
$\mathfrak{I}\bigl[K^{\textrm{det}}(\Omega)\bigr]$ (damping) are
plotted in Fig. (\ref{pic_K}). We do not consider radiative friction
here since $|K(\Omega)|\gg|2\Omega\,\Gamma(\Omega)|$;

\begin{align}
    X^{\textrm{det}}(\Omega)=&\frac{1}{2}\,Lh(\Omega)
    \left[1+\frac{K^{\textrm{det}}(\Omega)}{m\Omega^2-K^{\textrm{det}}(\Omega)}\,
    \frac{\sin\Omega\tau}{\Omega\tau}\right],\nonumber\\
    a^{\textrm{det}}_{\textrm{out}}(\Omega+\omega_0)=
    &-A_{\textrm{in}0}\,\frac{2\gamma}{\gamma-i\delta}\,\frac{1+2i\delta\tau}
    {1-R(1+2i\delta\tau)e^{2i\Omega\tau}}\nonumber\\
    &\times\,\frac{m\Omega^2}{m\Omega^2-K^{\textrm{det}}(\Omega)}\,
    ik_0Lh(\Omega)\,\frac{\sin\Omega\tau}{\Omega\tau}\,e^{i\Omega\tau}.
    \label{detuning_response}
\end{align}
Therefore, optical rigidity appears only when pumping frequency is
detuned from resonance.

In long-wave approximation ('lw') we obtain:
\begin{equation}
    K^{\textrm{det}}_{\textrm{lw}}(\Omega)=\frac{2\omega_0\mathcal{E}_{\textrm{FP}}}{L^2}\,
    \frac{\delta}{\delta^2+(\gamma-i\Omega)^2},
    \label{long_wave_rigidity}
\end{equation}
coinciding with the corresponding formula in Ref. \cite{rigidity_3}.
One can easily check that the ratio
$\mathfrak{R}\bigl[K^{\textrm{det}}(\Omega)\bigr]/(m\Omega^2)$ can
be made of the order of unity for the values of parameters planned
for Advanced LIGO ($L=4$ km, $\mathcal{E}_{\textrm{FP}}\leq 20$ J,
$m=40$ kg, $\omega_0/2\pi= 3\times 10^{14}$ Hz, $\gamma/2\pi\approx
1$ Hz, $\Omega/2\pi\approx\delta/2\pi=100$ Hz);
\begin{align}
    X^{\textrm{det}}_{\textrm{lw}}(\Omega)\approx
    &\frac{m\Omega^2}{m\Omega^2-K^{\textrm{det}}_{\textrm{lw}}(\Omega)}\,\frac{1}{2}\,Lh(\Omega),
    \label{long_wave_displacement}\\
    a^{\textrm{det}}_{\textrm{out, lw}}(\Omega+\omega_0)\approx
    &-A_{\textrm{in}0}\,\frac{\gamma/\tau}{\gamma-i\delta}\,
    \frac{1+2i\delta\tau}{\gamma-i(\delta+\Omega)}\nonumber\\
    &\times\frac{m\Omega^2}{m\Omega^2-K^{\textrm{det}}_{\textrm{lw}}(\Omega)}\,ik_0Lh(\Omega)
    \label{long_wave_response}.
\end{align}

Formulas (\ref{long_wave_rigidity}--\ref{long_wave_response}) are
applicable to consideration of the signal-recycling topology since
it is equivalent (with some modifications of parameters) to a single
detuned Fabry-Perot cavity \cite{rigidity_6}. Therefore, all the
major results of Refs.
\cite{rigidity_4,rigidity_5,rigidity_7,rigidity_8} which studied in
detail optical springs in SR topologies hold true for a single
cavity.

\subsection{Response function of the detuned cavity in the vicinity of FSR}
\begin{figure}[h]
\begin{center}
\includegraphics[scale=0.60]{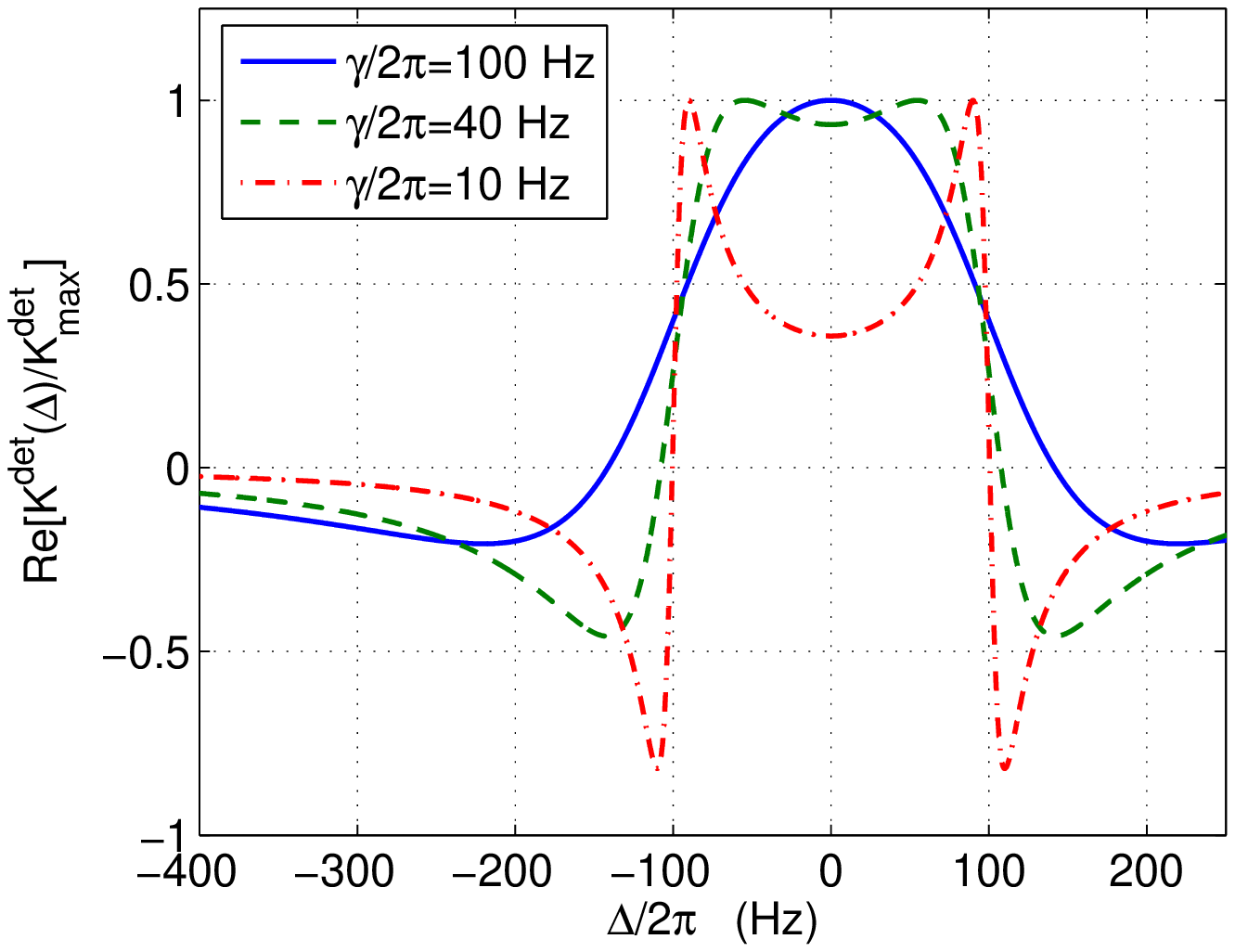}
\includegraphics[scale=0.60]{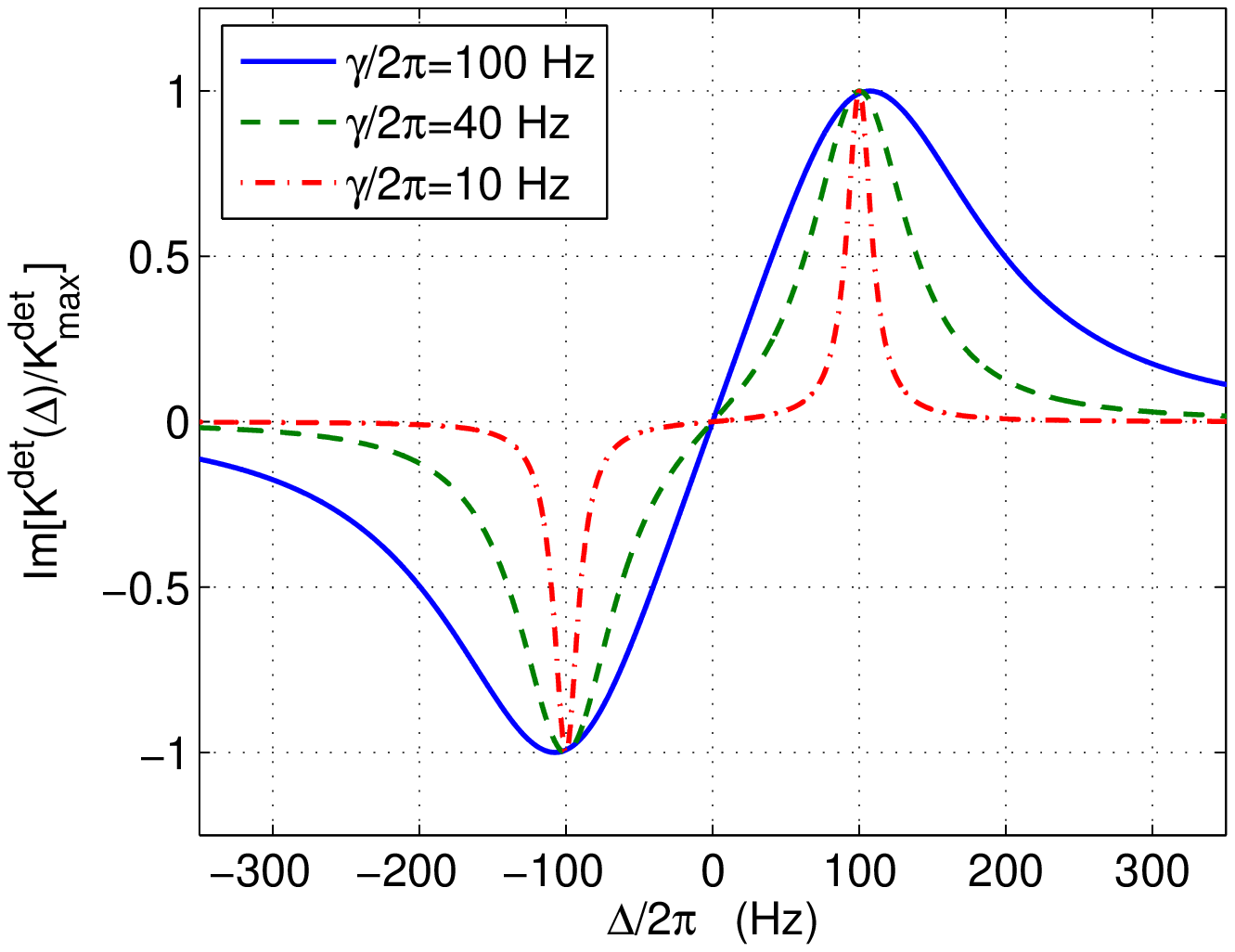}
\caption{Normalized rigidity (upper figure) and damping (lower
figure) as functions of detuning from the FSR. For visualization we
assumed $\delta/2\pi=100$ Hz.}\label{pic_K_doublet}
\end{center}
\end{figure}

To analyze the operation of GW detector near the FSR we turn to
formulas (\ref{detuning_rigidity}) and (\ref{detuning_response}).
Introducing notation $\Delta\equiv\Omega-\omega_{\textrm{fsr}}$,
$|\Delta|\ll\omega_{\textrm{fsr}}$ we rewrite these formulas in the
following way:
\begin{equation*}
    K^{\textrm{det}}(\Delta)=\frac{2\omega_0\mathcal{E}_{\textrm{FP}}}{L^2}\,
    \frac{\delta}{\delta^2+(\gamma-i\Delta)^2}.
\end{equation*}
We plotted $\mathfrak{R}\bigl[K^{\textrm{det}}(\Delta)\bigr]$ and
$\mathfrak{I}\bigl[K^{\textrm{det}}(\Delta)\bigr]$ in the vicinity
of $\Delta=0$ (or, equivalently, near
$\Omega=\omega_{\textrm{fsr}}$) in Fig. \ref{pic_K_doublet}. Note
that while decreasing the cavity half-bandwidth from
$\gamma\sim\delta$ to $\gamma\ll\delta$ the shape of
$\mathfrak{R}\bigl[K^{\textrm{det}}(\Delta)\bigr]$ changes from
``singlet'' to ``doublet''. Obviously, this takes place in the
vicinity of each $\Omega=n\omega_{\textrm{fsr}},\ n\in \mathbb{Z}$;
\begin{figure}[h]
\begin{center}
\includegraphics[scale=0.60]{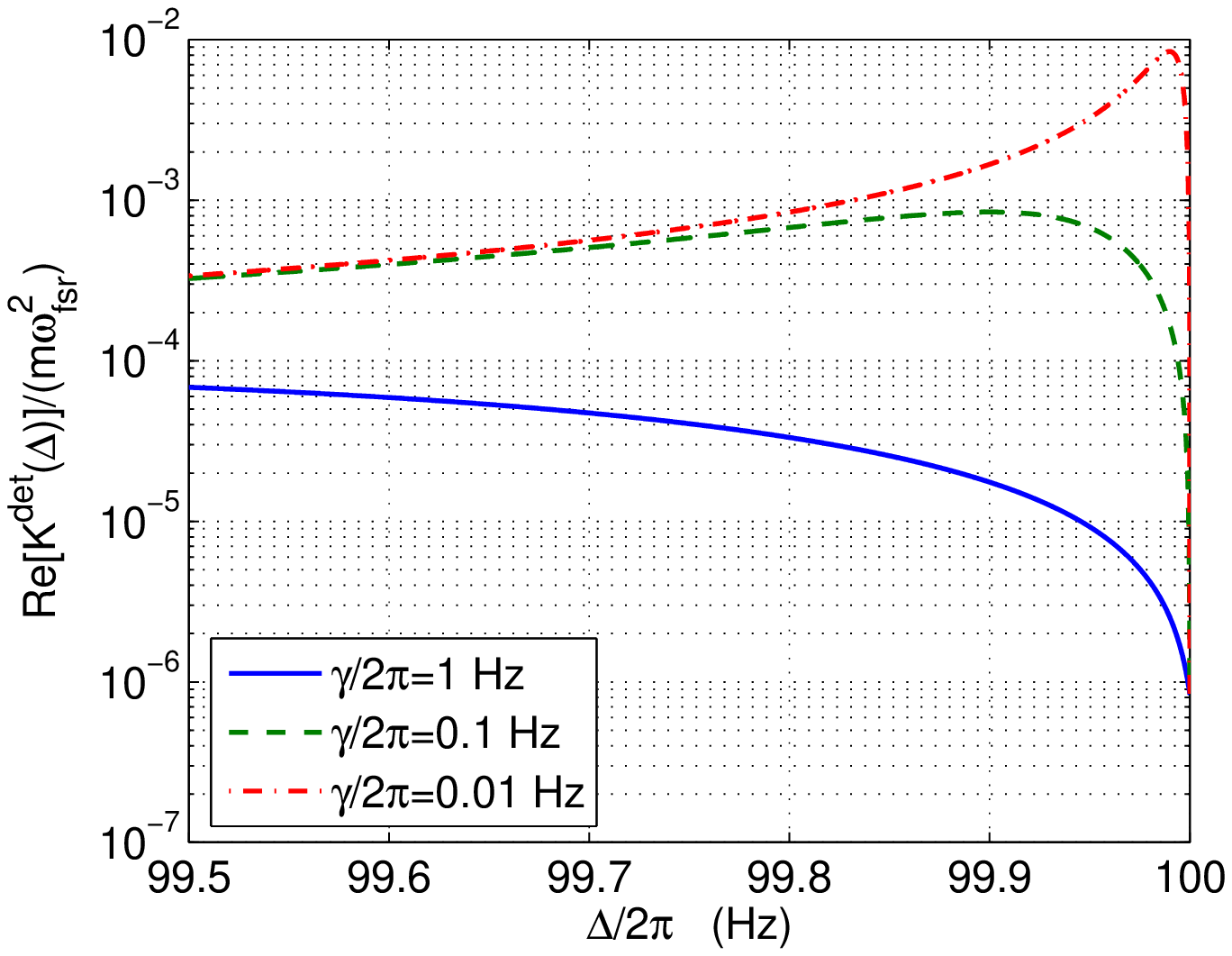}
\includegraphics[scale=0.60]{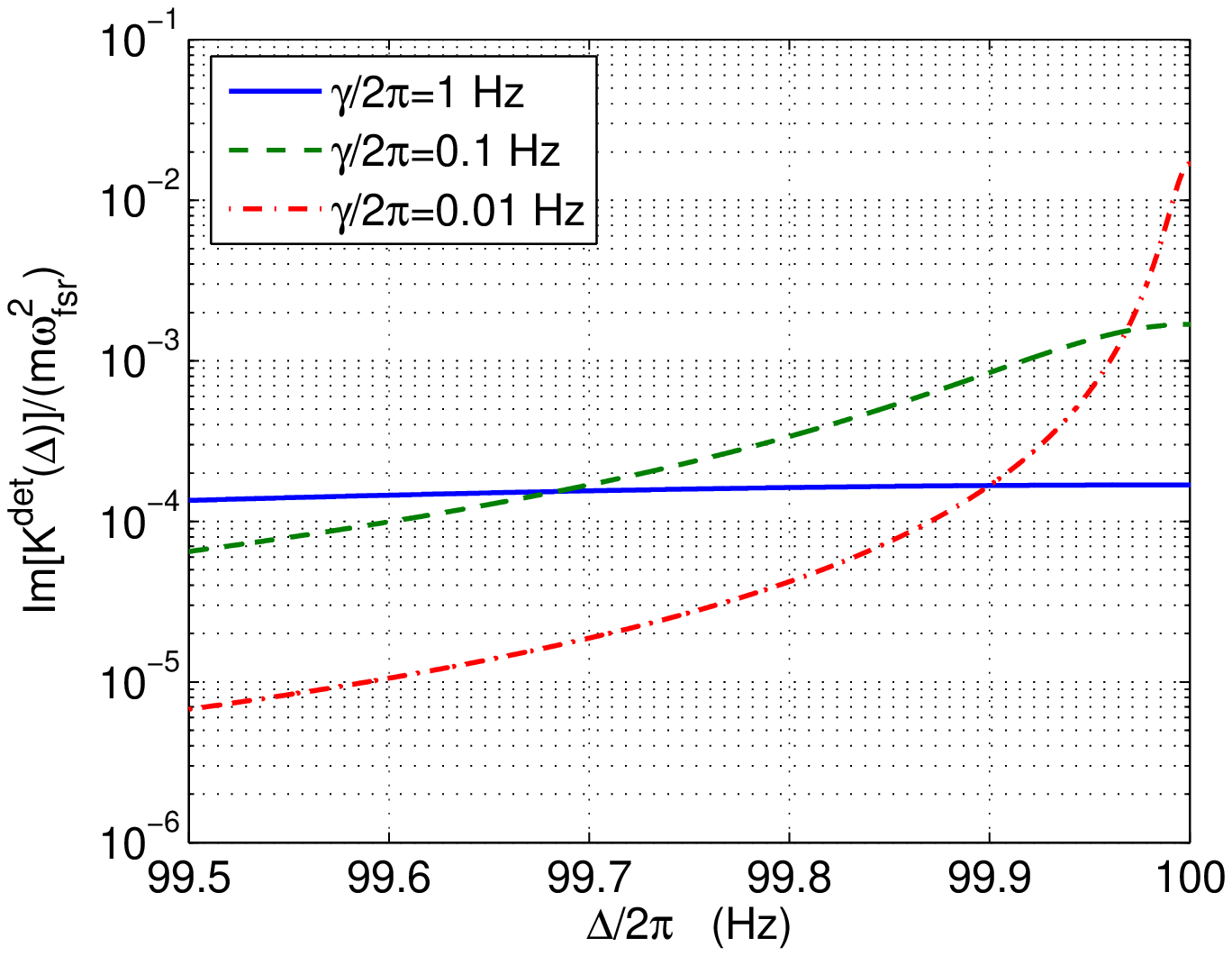}
\caption{Rigidity (upper figure) and damping (lower figure) near the
FSR compared to $m\omega_{\textrm{fsr}}^2$. For visualization we
assumed $L=4$ km, $\omega_0/2\pi=3\times 10^{14}$ Hz,
$\delta/2\pi=100$ Hz, $\mathcal{E}_{\textrm{FP}}=20$ J, $m=40$ kg
and $\omega_{\textrm{fsr}}/2\pi=37.5$ kHz.}\label{pic_K_FSR}
\end{center}
\end{figure}
\begin{align*}
    a^{\textrm{det}}_{\textrm{out}}&(\omega_{\textrm{fsr}}+\Delta+\omega_0)\\
    \approx &-A_{\textrm{in}0}\,\frac{\gamma/\tau}{\gamma-i\delta}\,
    \frac{1+2i\delta\tau}{\gamma-i(\delta+\Delta)}\\
    &\times\frac{m\omega_{\textrm{fsr}}^2}
    {m\omega_{\textrm{fsr}}^2-K^{\textrm{det}}(\Delta)}
    \,\frac{\Delta}{\omega_{\textrm{fsr}}}\,ik_0Lh(\omega_{\textrm{fsr}}+\Delta).
\end{align*}
The ratio
$\mathfrak{R}\bigl[K^{\textrm{det}}(\Delta)\bigr]/(m\omega_{\textrm{fsr}}^2)$
can be estimated as $\approx 10^{-4}$ (see Fig. \ref{pic_K_FSR}) for
$L=4$ km, $\omega_0/2\pi=3\times 10^{14}$ Hz, $\delta/2\pi=100$ Hz,
$\gamma/2\pi=1$ Hz, $\mathcal{E}_{\textrm{FP}}=20$ J, $m=40$ kg and
$\omega_{\textrm{fsr}}/2\pi=37.5$ kHz. Thus the implementation of
optical rigidity for the detection of signals near the FSR will
require the increase of circulating optical power, the decrease of
cavity half-bandwidth (see Fig. \ref{pic_K_FSR}) and the mirror
mass. With these requirements met it will be possible to perform the
resonant detection of high-frequency gravitational waves in the
vicinity of $37.5$ kHz. Note that while
$m\omega_{\textrm{fsr}}^2-\mathfrak{R}\bigl[K^{\textrm{det}}(\Delta)\bigr]$
can be made vanish for certain frequencies in principle, the
resonant gain will still be limited by
$m\omega_{\textrm{fsr}}^2/\mathfrak{I}\bigl[K^{\textrm{det}}(\Delta)\bigr]$.

\section{Conclusion}
In this paper we have analyzed the interaction of '+'-polarized
plane gravitational waves with a Fabry-Perot cavity. Calculations
have been performed in its input mirror's local Lorentz frame which
is best suited for consideration of non-gravitational forces acting
on test masses, such as the force of radiation pressure. We have
considered the general case of ratio $L/\lambda_{\textrm{gw}}$ ($L$
is the cavity length) for the purpose of high-frequency GW signals
detection when long-wave approximation is violated. Such
high-frequency gravitational waves can be emitted by the sources
predicted in modern astrophysical and cosmological theories, for
example, relic GW background.

We began from the solution of wave equation for running light wave
which is coupled to a weak plane gravitational wave and found that
in long-wave approximation this interaction leads to a small
correction of the order of $(x/\lambda_{\textrm{gw}})^2\,h$, where
$x$ is the distance traveled by light, in agreement with Ref.
\cite{thorne_clas_phys}. However if $x\sim\lambda_{\textrm{gw}}$
direct interaction between gravitational wave and light wave becomes
significant and is described by oscillating function. We applied
this result to consideration of the boundary problem for the light
wave in a Fabry-Perot cavity with the movable mirror. It was shown
that the response signal of such GW detector includes two terms: (i)
the phase shift of light due to reflection from the movable mirror
and (ii) the phase shift due to direct coupling of gravitational
wave to light wave. In our analysis we assumed that the end-mirror
of the cavity moves under the influence of (i) gravitational wave
and (ii) the force of light pressure, neglecting all other forces.
As a special case, by neglecting the resonant features of the
cavity, we examined the situation of single reflection when
radiation pressure is negligible. In this case test mass undergoes
geodesic motion only and, therefore, is inertial. TT gauge can be
applied then to the analysis of GW detector. This was performed in
Ref. \cite{thorne_clas_phys} and we found that our LL-based results
coincide with the TT-based results. This also confirmed the results
of Ref. \cite{local_observ}: in order to get the agreement between
the LL- and the TT-results direct interaction of the gravitational
wave with the light wave must be taken into account in the LL gauge
along with the displacement of test masses.

Next we considered the mirror equation of non-geodesic motion in the
field of gravitational wave and the force of light pressure.
Traditionally in literature this is done only in long-wave
approximation. We proposed rather general method of calculation of
radiation pressure when condition $L\ll\lambda_{\textrm{gw}}$ is
violated and showed that this calculation can be performed in flat
space-time without introducing a significant error. We obtained the
formulas for coefficients of ponderomotive optical rigidity and
radiative friction along with the general relativistic correction
describing the influence of direct 'GW+EMW' interaction on the force
of radiation pressure. This correction must be taken into account in
order to get correct formula for the response signal when long-wave
approximation is violated. Though formulas for optical rigidity,
radiative friction and 'GW+EMW' correction separately have been
presented in literature before
\cite{rigidity_3,rad_frict,rakh_phd_thesis}, nobody have obtained
them simultaneously using one unified method, as we have done in
this paper. The phenomenon of optical rigidity plays significant
role in operation of proposed schemes for advanced GW detectors and
our results may be of potential interest in consideration of the
regimes of high-frequency GW signals detection. One may also apply
our results to calculate the modification of the detector angular
pattern, including its magnetic component [35,36], due to radiation
pressure effects.

The possibility of parametric transitions between different resonant
curves of the cavity under the influence of gravitational waves with
frequencies $\omega_{\textrm{gw}}\approx n\omega_{\textrm{fsr}}$ was
discussed. This effect may find implementation in detection of the
high-frequency GW signals using optical modes with frequencies
$\omega_0\pm n\omega_{\textrm{fsr}}$ as signal waves instead of the
$\omega_0$-input mode.

In the end we considered two special cases of optical resonance and
small detuning from it. Excellent agreement with results of
\cite{rigidity_3,rad_frict} was found. We also estimated the deposit
of optical rigidity into the response signal near the FSR and showed
that in order to perform the resonant detection of GWs near that
frequency one should increase the circulating optical power and
decrease cavity half-bandwidth and the end-mirror mass.

\begin{acknowledgments}
I would like to thank S.P. Vyatchanin and  F.Ya. Khalili for useful
comments on the paper and fruitful discussions on physics of
interferometric GW antennas; V.I. Denisov for his comments and
advice on general relativistic aspects of my work; S.L. Danilishin
for careful reading of the manuscript and correction of errors in
the text. I also would like to express the sincerest gratitude to
Malik Rakhmanov who devoted much time and effort to give me detailed
comments and advice on the paper, pointed out my inaccuracies and
uncertainties and gave me some references I was not aware of. I also
thank him for very interesting and fruitful discussion on
high-frequency GWs and future prospects of their detection.

This work was supported by LIGO team from Caltech and in part by NSF
and Caltech grant PHY-0353775, by the Russian Agency of Industry and
Science, contracts No. 5178.2006.2 and 02.445.11.7423. This article
has been assigned LIGO doc. number LIGO-P060055-01-R.
\end{acknowledgments}

\newpage
\appendix
\section{Solution of the first order equation for running em wave}\label{1st_ord_sol}
In this Appendix we obtain the solution of the Cauchy problem
(\ref{1st_order_equation}--\ref{initial_conditions}).

It is convenient to go to spectral domain. Applying the theorem of
convolution to the right side of Eq. (\ref{1st_order_equation}), we
get the following first-order equation in spectral domain:
\begin{subequations}
\begin{align}
    \frac{\partial^2A^{(1)}}{\partial x^2}+\frac{\Omega^2}{c^2}&A^{(1)}\nonumber\\
    =\frac{1}{2}k_0^2x^2&\biggl[
    A_{+0}e^{ik_0x}\,\frac{(\Omega-\omega_0)^2}{c^2}\,h(\Omega-\omega_0)\nonumber\\
    &+A_{+0}^*e^{-ik_0x}\,\frac{(\Omega+\omega_0)^2}{c^2}\,h(\Omega+\omega_0)\nonumber\\
    &+A_{-0}e^{-ik_0x}\,\frac{(\Omega-\omega_0)^2}{c^2}\,h(\Omega-\omega_0)\nonumber\\
    &+A_{-0}^*e^{ik_0x}\,\frac{(\Omega+\omega_0)^2}{c^2}\,h(\Omega+\omega_0)\biggr],
    \label{appendix_A_1st_order_spectral_eq}
\end{align}
where
\begin{align}
    &A^{(1)}(x,\Omega)=A_+^{(1)}(x,\Omega)+A_-^{(1)}(x,\Omega),
    \label{appendix_A_sum}\\
    &A_+^{(1)}(0,\Omega)=A_-^{(1)}(0,\Omega)=0.
    \label{appendix_A_initial_conditions}
\end{align}
\end{subequations}
are equivalent to formulas (\ref{sum}) and
(\ref{initial_conditions}) correspondingly. For convenience we
introduce the following notations:
\begin{align*}
    &A^{(1)}_+(x,\Omega)\equiv A,\qquad k\equiv\frac{\Omega}{c},\\
    &\mathcal{B}_1\equiv\frac{1}{2}\,k_0^2A_{+0}\,\frac{(\Omega-\omega_0)^2}{c^2}\,
    h(\Omega-\omega_0)\\
    &\mathcal{B}_2\equiv\frac{1}{2}\,k_0^2A_{+0}^*\,\frac{(\Omega+\omega_0)^2}{c^2}\,
    h(\Omega+\omega_0)\\
    &\mathcal{B}_3\equiv\frac{1}{2}\,k_0^2A_{-0}\,\frac{(\Omega-\omega_0)^2}{c^2}\,
    h(\Omega-\omega_0)\\
    &\mathcal{B}_4\equiv\frac{1}{2}\,k_0^2A_{-0}^*\,\frac{(\Omega+\omega_0)^2}{c^2}\,
    h(\Omega+\omega_0).
\end{align*}
Eq. (\ref{appendix_A_1st_order_spectral_eq}) rewritten in these
notations is:
\begin{equation}
    \frac{d^2A}{dx^2}+k^2A=(\mathcal{B}_1+\mathcal{B}_4)x^2e^{ik_0x}+
    (\mathcal{B}_2+\mathcal{B}_3)x^2e^{-ik_0x}.
    \label{appendix_A_1st_order_eq_notations}
\end{equation}
Note that spectral frequency $\Omega$ (or $k=\Omega/c$) plays the
role of parameter here. General solution of homogeneous equation
$d^2A/dx^2+k^2A=0$ is well-known: $A(x)=C_1e^{ikx}+C_2e^{-ikx}$. We
shall obtain the particular solution of heterogeneous Eq.
(\ref{appendix_A_1st_order_eq_notations}) using the method of
variation of constants (see any ODE handbook), assuming
\begin{equation}
    A(x)=C_1(x)e^{ikx}+C_2(x)e^{-ikx}.
    \label{appendix_A_variation_of_constants}
\end{equation}
Following the general rule of this method we must solve the
following set of equations:
\begin{align*}
    \frac{dC_1}{dx}\,e^{ikx}+\frac{dC_2}{dx}\,e^{-ikx}=&0,\\
    ik\,\frac{dC_1}{dx}\,e^{ikx}-ik\,\frac{dC_2}{dx}\,e^{-ikx}=&
    (\mathcal{B}_1+\mathcal{B}_4)x^2e^{ik_0x}\\
    &+(\mathcal{B}_2+\mathcal{B}_3)x^2e^{-ik_0x}.
\end{align*}
It is straightforward to verify that solution is
\begin{align*}
    C_1(x)=&\frac{\mathcal{B}_1+\mathcal{B}_4}{2ik}\,e^{i(k_0-k)x}\\
    &\times \left(\frac{x^2}{i(k_0-k)}+
    \frac{2x}{(k_0-k)^2}-\frac{2}{i(k_0-k)^3}\right)\\
    +&\frac{\mathcal{B}_2+\mathcal{B}_3}{2ik}\,e^{-i(k_0+k)x}\\
    &\times\left(-\frac{x^2}{i(k_0+k)}+
    \frac{2x}{(k_0+k)^2}+\frac{2}{i(k_0+k)^3}\right)\\
    +&C_{10}.
\end{align*}
and
\begin{align*}
    C_2(x)=&-\frac{\mathcal{B}_1+\mathcal{B}_4}{2ik}\,e^{i(k_0+k)x}\\
    &\times\left(\frac{x^2}{i(k_0+k)}+
    \frac{2x}{(k_0+k)^2}-\frac{2}{i(k_0+k)^3}\right)\\
    &-\frac{\mathcal{B}_2+\mathcal{B}_3}{2ik}\,e^{-i(k_0-k)x}\\
    &\times\left(-\frac{x^2}{i(k_0-k)}+
    \frac{2x}{(k_0-k)^2}+\frac{2}{i(k_0-k)^3}\right)\\
    +&C_{20}.
\end{align*}

Substituting these solutions for $C_1(x)$ and $C_2(x)$ into formula
(\ref{appendix_A_variation_of_constants}) and keeping in mind our
notations and formulas (\ref{xi}--\ref{zeta},\ \ref{appendix_A_sum})
we obtain the general solution for ``positive'' and ``negative''
waves:
\begin{align}
    A^{(1)}_\pm(x,\Omega)=C_{\pm0}e^{\pm i\frac{\Omega}{c}x}\nonumber\\+
    \frac{1}{2}\,&A_{\pm0}h(\Omega-\omega_0)e^{\pm ik_0x}\nonumber\\
    \times&\Bigl[x^2\xi(\Omega)\mp ix\eta(\Omega)-\zeta(\Omega)\Bigr]\nonumber\\
    +\frac{1}{2}\,&A_{\pm0}^*h(\Omega+\omega_0)e^{\mp ik_0x}\nonumber\\
    \times&\Bigl[x^2\xi(-\Omega)\pm ix\eta(-\Omega)-\zeta(-\Omega)\Bigr],
    \label{appendix_A_1st_order_solution}
\end{align}
Here we redefined for convenience constants $C_{10}$ and $C_{20}$ as
$C_{+0}$ and $C_{-0}$ to describe separately the additions to
``positive'' and ``negative'' waves. Constants $C_{\pm0}$ are
obtained from initial conditions
(\ref{appendix_A_initial_conditions}):
\begin{equation*}
    C_{\pm0}=\frac{1}{2}\,A_{\pm0}h(\Omega-\omega_0)\zeta(\Omega)+
    \frac{1}{2}\,A_{\pm0}^*h(\Omega+\omega_0)\zeta(-\Omega).
\end{equation*}
Substituting them into formula (\ref{appendix_A_1st_order_solution})
and performing inverse Fourier transformation we finally obtain the
first order solution in time domain (\ref{1st_order_solution}).

\section{General relativistic corrections to the force of light pressure}\label{GR_corrections}
In this Appendix we shall prove the statement made in Sec.
\ref{eq_of_mot}: the 4-force of light pressure
$\mathcal{F}^\mu_{\textrm{em}}$ can be calculated by means of
classical flat space-time electrodynamics without introducing a
significant error.

Whatever the general relativistic rule for calculation of the light
pressure force in any case it will be proportional to: (i) the
components of energy-stress tensor $T^{\alpha\beta}$ of the light
field inside the cavity averaged over period of optical oscillations
$2\pi/\omega_0$, (ii) cross-section of the light beam $S$ and (iii)
components of metric tensor (\ref{metric_tensor}):
\begin{equation*}
    \mathcal{F}^\mu_{\textrm{em}}\propto g_{\lambda\nu}T^{\alpha\beta}S.
\end{equation*}
The energy-stress tensor of electromagnetic field in curved space
time is \cite{field_theory}:
\begin{equation*}
    T_{\mu\nu}=\frac{1}{4\pi}\left(F_{\mu\alpha}F_{\nu\beta}g^{\alpha\beta}-
    \frac{1}{4}F_{\alpha\beta}F^{\alpha\beta}g_{\mu\nu}\right),
\end{equation*}
where $F_{\alpha\beta}$ is the tensor of electromagnetic field
$A^{\gamma}$ inside the cavity:
\begin{align*}
    F_{\alpha\beta}=&\nabla_\alpha A_\beta-\nabla_\beta A_\alpha=
    g_{\beta\gamma}\nabla_\alpha A^\gamma-g_{\alpha\gamma}\nabla_\beta
    A^\gamma\\
    =&g_{\beta\gamma}\left(\frac{\partial A^\gamma}{\partial x^\alpha}+
    \Gamma^\gamma_{\nu\alpha}A^\nu\right)-
    g_{\alpha\gamma}\left(\frac{\partial A^\gamma}{\partial x^\beta}+
    \Gamma^\gamma_{\nu\beta}A^\nu\right).
\end{align*}
Here $\nabla_\alpha$ is the covariant derivative with respect to the
connection of space-time $\Gamma^{\mu}_{\nu\lambda}$:
\begin{equation*}
    \Gamma^{\mu}_{\nu\lambda}=\frac{1}{2}g^{\mu\alpha}
    \left(\frac{\partial g_{\alpha\nu}}{\partial x^\lambda}+
    \frac{\partial g_{\alpha\lambda}}{\partial x^\nu}-
    \frac{\partial g_{\nu\lambda}}{\partial x^\alpha}\right).
\end{equation*}

We shall now examine the orders of the values included in
$F_{\alpha\beta}$. To do this we shall omit all numerical constants
of the order of unity and all terms of the second and greater orders
in $h$. Besides we shall consider only the case of monochromatic GW
with frequency $\omega_{\textrm{gw}}$ since generalization is
trivial. Remind that all functions are evaluated at point $x=L$.
Returning to Sec. \ref{intro} we conclude that
\begin{equation*}
    g_{\mu\nu}\sim 1+\frac{L^2}{c^2}\,\ddot{h},\qquad
    \Gamma^{\mu}_{\nu\lambda}\sim
    \frac{L^2}{c^2}\,\frac{\omega_{\textrm{gw}}}{c}\,\ddot{h}+
    \frac{L}{c^2}\,\ddot{h}.
\end{equation*}

Remind that (see formula (\ref{A_inside}) electromagnetic field
$A^\gamma(x,t)$ inside the cavity is a sum of ``positive'' wave
$A_+(x,t)$ and ``negative'' wave $A_-(x,t)$. These waves include
``big'' components with amplitudes $A_{\pm0}$ and ``small''
components proportional to $A_{\pm0}h$ (we are not interested here
in the term proportional to $X$). It is necessary to mention that
these ``small'' terms are included in the wavefunctions of
electromagnetic field and, in general, the gravitational nature of
these small corrections is of no importance to us here. Therefore,
we shall not consider them as general relativistic corrections
further, since we are interested only in GR corrections to the
\textit{methods} of flat space-time electrodynamics but not the
corrections included in $A^\gamma(x,t)$ itself. We estimate the
order of the derivative of EMW field as
\begin{equation*}
    \frac{\partial A}{\partial x}\biggr|_L\sim k_0A_0+k_0A_0H(L,t),
\end{equation*}
where $H(L,t)\sim h$ is the mentioned small correction to the EMW
field inside the cavity, containing GW wavefunction. Thus we can
estimate the order of $F_{\alpha\beta}$ in the following way:
\begin{equation*}
    F_{\alpha\beta}
    \sim k_0A_0\left(1+H(L,t)+\frac{L^2}{c^2}\,\ddot{h}+\frac{L^2}{c^2}\,
    \frac{\omega_{\textrm{gw}}}{\omega_0}\,\ddot{h}+
    \frac{L}{\omega_0c}\,\ddot{h}\right).
\end{equation*}
Here $A_0$ has the order of magnitude of the electromagnetic wave's
amplitude inside the cavity. The light pressure 4-force
$\mathcal{F}^\mu_{\textrm{em}}$ is proportional to
$F_{\alpha\beta}$, ``squared'' (remind that we neglect all numerical
coefficients), averaged over period of optical oscillations
$2\pi/\omega_0$ (terms proportional to $e^{2i\omega_0t}$ disappear)
and multiplied by $S$:
\begin{align*}
    \mathcal{F}^\mu_{\textrm{em}}\sim &W_{\textrm{FP}}S\\
    &\times\left(1+H(L,t)+\frac{L^2}{c^2}\,\ddot{h}+\frac{L^2}{c^2}\,
    \frac{\omega_{\textrm{gw}}}{\omega_0}\,\ddot{h}+\frac{L}{\omega_0c}\,\ddot{h}\right),
\end{align*}
where $W_{\textrm{FP}}\sim k_0^2A_0^2$ is proportional to the energy
density of light wave inside the cavity.

Let us now return to the mirror's equation of motion:
\begin{equation*}
    \frac{d^2x^\mu}{ds^2}+\Gamma^{\mu}_{\nu\lambda}\frac{dx^\nu}{ds}
    \frac{dx^\lambda}{ds}=\frac{1}{m}\,\mathcal{F}^\mu_{\textrm{em}}.
\end{equation*}
We rewrite this equation in non-relativistic limit for
$x$-component, expanding the force of light pressure into the zeroth
and the first order in $h$ terms:
\begin{equation*}
    \frac{d^2x}{dt^2}-\frac{1}{2}x\ddot{h}(t)=
    \frac{1}{m}\left(F^{(0)}+F^{(1)}_{\textrm{FP}}+
    F^{(1)}_{\textrm{GR}}\right),
\end{equation*}
where
\begin{equation*}
    F^{(0)}\sim W_{\textrm{FP}}S,
\end{equation*}
is the constant force of light pressure;
\begin{equation*}
    F^{(1)}_{\textrm{FP}}\sim W_{\textrm{FP}}SH(L,t),
\end{equation*}
is the part of the force proportional to the phase shift of light
due to its direct interaction with GW;
\begin{equation*}
    F^{(1)}_{\textrm{GR}}\sim W_{\textrm{FP}}S
    \left(\frac{L^2}{c^2}\,\ddot{h}+
    \frac{L^2}{c^2}\,\frac{\omega_{\textrm{gw}}}{\omega_0}\,\ddot{h}+
    \frac{L}{\omega_0c}\,\ddot{h}\right),
\end{equation*}
is the GR correction to the methods of flat space-time
electrodynamics. Our problem is to prove that it can be omitted when
calculating the force of light pressure. We use the method of
successive approximations to solve the equation of motion. The
zeroth order force is constant, therefore we should redefine the
length $L$ in all formulas, assuming that in real experiment
constant force is compensated somehow. The first order equation is
then
\begin{equation*}
    \frac{d^2X}{dt^2}-\frac{1}{2}\,L\ddot{h}(t)=
    \frac{1}{m}\,F^{(1)}_{\textrm{FP}}+\frac{1}{m}\,F^{(1)}_{\textrm{GR}}.
\end{equation*}
To estimate the orders of all the terms included in this equation we
turn to the spectral domain:
\begin{equation*}
    -\Omega^2X(\Omega)+\frac{1}{2}\,L\Omega^2h(\Omega)=
    \frac{1}{m}\,F^{(1)}_{\textrm{FP}}(\Omega)+\frac{1}{m}\,F^{(1)}_{\textrm{GR}}(\Omega).
\end{equation*}
Its solution is:
\begin{align}
    X&(\Omega)\sim
    Lh(\Omega)+\frac{W_{\textrm{FP}}S}{m\Omega^2}\,H(L,\Omega)\nonumber\\
    &+\frac{W_{\textrm{FP}}S}{m\Omega^2}\left(\frac{L^2}{c^2}\,\omega_{\textrm{gw}}^2+
    \frac{L^2}{c^2}\,\frac{\omega_{\textrm{gw}}}{\omega_0}\,
    \omega_{\textrm{gw}}^2+\frac{L}{\omega_0c}\,\omega_{\textrm{gw}}^2\right)h(\Omega).
    \label{corrections_to_x}
\end{align}
We can replace spectral frequency $\Omega$ with
$\omega_{\textrm{gw}}$ since $h(\Omega)\sim
h_0\delta(\Omega\pm\omega_{\textrm{gw}})$. Then
\begin{align*}
    X(\Omega)\sim &Lh(\Omega)+\frac{W_{\textrm{FP}}S}{m\Omega^2}H(L,\Omega)\\
    &+\frac{{\mathcal{E}_{\textrm{FP}}}}{mc^2}\left(1+\frac{\lambda_0}{\lambda_{\textrm{gw}}}\,
    +\frac{\lambda_0}{L}\right)Lh(\Omega),
\end{align*}
where $\mathcal{E}_{\textrm{FP}}=W_{\textrm{FP}}SL$ is the full
energy of light wave inside the cavity. We can neglect the third
term in formula (\ref{corrections_to_x}) completely since
$\mathcal{E}_{\textrm{FP}}/mc^2\approx 10^{-17}$
($\mathcal{E}_{\textrm{FP}}=20$ J and $m=40$ kg for Advanced LIGO
project), $\lambda_0/\lambda_{\textrm{gw}}\sim 10^{-13}\div
10^{-10}$ and $\lambda_0/L\approx 10^{-9}$.

Now looking back, we see that the term we have just omitted
corresponds to $F^{(1)}_{\textrm{GR}}$ --- the statement required to
be proved.

\section{Calculation of the force of light pressure in an explicit form}\label{force}
In this Appendix we calculate explicitly the force of light pressure
\begin{equation*}
    F_x(X,\dot{X},t)=
    \frac{S}{8\pi}\left[\frac{1}{c^2}\left(\frac{\partial A}{\partial t}\right)^2+
    \left(\frac{\partial A}{\partial x}\right)^2\right]^{(1)}_{x=L+X(t)},
\end{equation*}
which is the right side of Eq. (\ref{equation_of_motion}) (for
simplicity we use notation $F_x$ instead of $F_x^{(1)}$).

\subsection{General formula\label{force_gen_formula}}
First we calculate the force of light pressure in time domain.
Summing up the derivatives of $A_+(x,t)$ and $A_-(x,t)$ (see formula
(\ref{A_inside})), raising the sum to the second power, neglecting
the terms containing $e^{2i\omega_0t}$ and keeping only the first
order in $h$ terms, we obtain:
\begin{align*}
    \left.\left(\frac{\partial A}{\partial x}\right)^2\right|_{L+X(t)}
    =2ik_0\biggl[
    &A_{+0}\left(\frac{\partial a^*_+}{\partial x}-ik_0a_+^*\right)\\
    +&A_{+0}\left(\frac{\partial a^*_-}{\partial x}+ik_0a_-^*\right)e^{2ik_0L}\\
    -&A_{-0}\left(\frac{\partial a^*_-}{\partial x}+ik_0a_-^*\right)\\
    -&A_{-0}\left(\frac{\partial a^*_+}{\partial x}-ik_0a_+^*\right)e^{-2ik_0L}
    \biggl]_L\\
    +\textrm{c.c.}&
\end{align*}
Substituting formula $A_{+0}=-A_{-0}e^{-2i\omega_0\tau}$ here we
obtain:
\begin{align}
    \left.\left(\frac{\partial A}{\partial x}\right)^2\right|_{L+X(t)}
    =&4ik_0A_{+0}\left(\frac{\partial a^*_+}{\partial x}-ik_0a_+^*\right)_L\nonumber\\
    -&4ik_0A_{-0}\left(\frac{\partial a^*_-}{\partial x}+ik_0a_-^*\right)_L
    +\textrm{c.c.}
    \label{dA_dx_squared}
\end{align}

Similar calculations lead to
\begin{equation*}
    \left.\frac{1}{c^2}\left(\frac{\partial A}{\partial t}\right)^2
    \right|_{L+X(t)}=0.
\end{equation*}
This is not surprising since $E_z(L+X)=-(\dot{X}/c)H_y(L+X)$ (see
formula (\ref{electric_field_movable})) and being squared leads to
$X^2$. Therefore, $E_z^2$ equals to zero in our linear
approximation.

Substituting formula (\ref{a+-_forier_int}) into
(\ref{dA_dx_squared}) we obtain the final formula for $F_x$:
\begin{equation*}
    F_x(X,\dot{X},t)=\int_{-\infty}^{+\infty}F_x(\Omega)e^{-i\Omega t}\,\frac{d\Omega}{2\pi},
\end{equation*}
\begin{align}
    F_x(\Omega)=\frac{Sk_0}{2\pi}\biggl[
    &A_{+0}^*a_+(\omega_0+\Omega)\,\frac{\omega_0+\Omega}{c}\,e^{i\Omega\tau}\nonumber\\
    +&A_{+0}a_+^*(\omega_0-\Omega)\,\frac{\omega_0-\Omega}{c}\,e^{i\Omega\tau}\nonumber\\
    +&A_{-0}^*a_-(\omega_0+\Omega)\,\frac{\omega_0+\Omega}{c}\,e^{-i\Omega\tau}\nonumber\\
    +&A_{-0}a_-^*(\omega_0-\Omega)\,\frac{\omega_0-\Omega}{c}\,e^{-i\Omega\tau}\biggr].
    \label{light_pressure}
\end{align}
Now we substitute explicit formulas (\ref{a_plus}) and
(\ref{a_minus}) here and split $F_x(\Omega)$ into several parts
according to their physical sense:
\begin{equation*}
    F_x(\Omega)=F_{\textrm{fluct}}(\Omega)+F_{\textrm{pm}}(\Omega)+F_{\textrm{gw+emw}}(\Omega).
\end{equation*}

Note that we have not taken into account the terms proportional to
$A_{\pm0}g_\pm(x,t)$ while calculating the force of light pressure.
It is straightforward to verify, using similar calculations, that
the corresponding correction can be estimated as $\delta
F_x(\Omega)\approx SW_{\textrm{FP}}(\Omega\tau)^2h(\Omega)$,
resulting in inaccuracy $\delta X(\Omega)\approx
(\mathcal{E}_{\textrm{FP}}/mc^2)Lh(\Omega)\approx
10^{-17}Lh(\Omega)$.

\subsection{Fluctuating force}\label{force_fluct}
Here we obtain formula for the fluctuating part
$F_{\textrm{fluct}}(\Omega)$ of the radiation pressure arising due
to fluctuations $a_{\textrm{in}}(x,t)$ of the input light wave:
\begin{align*}
    F_{\textrm{fluct}}(\Omega)\\
    =\frac{Sk_0}{2\pi}\Biggl[
    &a_{\textrm{in}}(\omega_0+\Omega)\,
    \frac{A_{+0}^*Te^{i\Omega\tau}}{1-Re^{2i(\omega_0+\Omega)\tau}}\,
    \frac{\omega_0+\Omega}{c}\\
    +&a_{\textrm{in}}^*(\omega_0-\Omega)\,
    \frac{A_{+0}Te^{i\Omega\tau}}{1-Re^{-2i(\omega_0-\Omega)\tau}}\,
    \frac{\omega_0-\Omega}{c}\\
    -&a_{\textrm{in}}(\omega_0+\Omega)\,
    \frac{A_{-0}^*Te^{2i(\omega_0+\Omega)\tau}e^{-i\Omega\tau}}{1-Re^{2i(\omega_0+\Omega)\tau}}\,
    \frac{\omega_0+\Omega}{c}\\
    -&a_{\textrm{in}}^*(\omega_0-\Omega)\,
    \frac{A_{-0}Te^{-2i(\omega_0-\Omega)\tau}e^{-i\Omega\tau}}{1-Re^{-2i(\omega_0-\Omega)\tau}}\,
    \frac{\omega_0-\Omega}{c}\Biggr].
\end{align*}
Using formulas (\ref{A+_Ain}) and (\ref{A-_Ain}) we obtain:
\begin{align*}
    F_{\textrm{fluct}}(\Omega)&\\
    =\frac{Sk_0}{\pi c}\Biggl[
    &\frac{T^2A^*_{\textrm{in}0}}{1-Re^{-2i\omega_0\tau}}\,
    \frac{(\omega_0+\Omega)e^{i\Omega\tau}}{1-Re^{2i(\omega_0+\Omega)\tau}}\,
    a_{\textrm{in}}(\omega_0+\Omega)\\
    &+\frac{T^2A_{\textrm{in}0}}{1-Re^{2i\omega_0\tau}}\,
    \frac{(\omega_0-\Omega)e^{i\Omega\tau}}{1-Re^{-2i(\omega_0-\Omega)\tau}}\,
    a^*_{\textrm{in}}(\omega_0-\Omega)\Biggr].
\end{align*}

\subsection{Ponderomotive force}\label{force_optmech_couple}
Here we obtain the formula for the coefficient $\mathcal{K}(\Omega)$
and corresponding ponderomotive force
$F_{\textrm{pm}}(\Omega)=-\mathcal{K}(\Omega)X(\Omega)$ (see
formulas (\ref{optical_rigidity}) and (\ref{radiation_friction})).
Ponderomotive force is the sum of all the terms in formula
(\ref{light_pressure}) that include phase shift
(\ref{phase_shift_mir}) arising due to mirror's displacement $X$
(see formulas (\ref{a_plus}) and (\ref{a_minus})):
\begin{align*}
    F_{\textrm{pm}}(\Omega)\\
    =\frac{SW_{\textrm{FP}}}{2k_0}&\Biggl[
    \frac{Re^{2i\omega_0\tau}e^{i\Omega\tau}}{1-Re^{2i(\omega_0+\Omega)\tau}}\,
    +\frac{e^{-i\Omega\tau}}{1-Re^{2i(\omega_0+\Omega)\tau}}\Biggr]\\
    &\times\frac{\omega_0+\Omega}{c}\,i\,\delta\Psi_{\textrm{mir}}(\Omega)\\
    -\frac{SW_{\textrm{FP}}}{2k_0}&\Biggl[
    \frac{Re^{-2i\omega_0\tau}e^{i\Omega\tau}}{1-Re^{-2i(\omega_0-\Omega)\tau}}\,
    +\frac{e^{-i\Omega\tau}}{1-Re^{-2i(\omega_0-\Omega)\tau}}\Biggr]\\
    &\times\frac{\omega_0-\Omega}{c}\,i\,\delta\Psi^*_{\textrm{mir}}(-\Omega).
\end{align*}
Substituting formula (\ref{phase_shift_mir}) here we obtain the
following formula for $\mathcal{K}(\Omega)$:
\begin{align}
    \mathcal{K}(\Omega)
    =-i\,SW_{\textrm{FP}}\biggl[
    &\frac{1+Re^{2i(\omega_0+\Omega)\tau}}{1-Re^{2i(\omega_0+\Omega)\tau}}
    \frac{\omega_0+\Omega}{c}\nonumber\\
    &-\frac{1+Re^{-2i(\omega_0-\Omega)\tau}}{1-Re^{-2i(\omega_0-\Omega)\tau}}
    \frac{\omega_0-\Omega}{c}
    \biggr]
    \label{appendix_C_rigidity}
\end{align}
It will be convenient for the future prospects to divide
$\mathcal{K}(\Omega)$ into two parts:
$\mathcal{K}(\Omega)=K(\Omega)-2i\Omega\,\Gamma(\Omega)$ using the
$(\omega_0\pm\Omega)/c$ multipliers. The first part $K(\Omega)$ is
proportional to $\omega_0/c$ and is called the coefficient of
optical rigidity:
\begin{equation*}
    K(\Omega)=\frac{4k_0SW_{\textrm{FP}}Re^{2i\Omega\tau}\sin2\omega_0\tau}
    {1-2Re^{2i\Omega\tau}\cos2\omega_0\tau+R^2e^{4i\Omega\tau}},
\end{equation*}
The second part $-2i\Omega\,\Gamma(\Omega)$ is proportional to
$\Omega/c$ where $\Gamma(\Omega)$ is called the coefficient of
radiative friction:
\begin{align*}
    \Gamma(\Omega)=\frac{SW_{\textrm{FP}}}{c}\,\frac{1-R^2e^{4i\Omega\tau}}
    {1-2Re^{2i\Omega\tau}\cos2\omega_0\tau+R^2e^{4i\Omega\tau}}.
\end{align*}
Therefore, we secure full analogy with linear harmonic oscillator's
mechanical impedance. Obviously,
$|2\Omega\Gamma|\sim(\Omega/\omega_0)|K|$.

\subsection{GW+EMW correction}\label{force_gw_emw_correction}
Here we obtain formula (\ref{gw_emw_correction}) for the 'GW+EMW'
correction to the force of light pressure. It arises due to direct
interaction of gravitational wave with the light wave inside the
cavity and becomes significant when length of the cavity becomes
comparable with gravitational wavelength. Therefore, this part of
(\ref{light_pressure}) is composed of the terms that include phase
shift (\ref{phase_shift_gw_emw}) arising due to direct coupling
'GW+EMW':
\begin{align*}
    F_{\textrm{gw+emw}}(\Omega)\\
    =\frac{SW_{\textrm{FP}}}{2k_0}&\Biggl[
    \frac{Re^{2i\omega_0\tau}e^{i\Omega\tau}}{1-Re^{2i(\omega_0+\Omega)\tau}}\,
    +\frac{e^{-i\Omega\tau}}{1-Re^{2i(\omega_0+\Omega)\tau}}\Biggr]\\
    &\times\frac{\omega_0+\Omega}{c}\,i\,\delta\Psi_{\textrm{gw+emw}}(\Omega)\\
    -\frac{SW_{\textrm{FP}}}{2k_0}&\Biggl[
    \frac{Re^{-2i\omega_0\tau}e^{i\Omega\tau}}{1-Re^{-2i(\omega_0-\Omega)\tau}}\,
    +\frac{e^{-i\Omega\tau}}{1-Re^{-2i(\omega_0-\Omega)\tau}}\Biggr]\\
    &\times\frac{\omega_0-\Omega}{c}\,i\,\delta\Psi^*_{\textrm{gw+emw}}(-\Omega).
\end{align*}
Substituting formula (\ref{direct_interaction}) here we obtain:
\begin{align*}
    F_{\textrm{gw+emw}}(\Omega)
    =-&\frac{1}{2}\,Lh(\Omega)\left(1-\frac{\sin\Omega\tau}{\Omega\tau}\right)
    i\,SW_{\textrm{FP}}\\
    \times\biggl[
    &\frac{1+Re^{2i(\omega_0+\Omega)\tau}}{1-Re^{2i(\omega_0+\Omega)\tau}}
    \frac{\omega_0+\Omega}{c}\\
    &-\frac{1+Re^{-2i(\omega_0-\Omega)\tau}}{1-Re^{-2i(\omega_0-\Omega)\tau}}
    \frac{\omega_0-\Omega}{c}
    \biggr]
\end{align*}
Taking into account formula (\ref{appendix_C_rigidity}) we finally
obtain the following formula for the 'GW+EMW' correction:
\begin{align*}
    F_{\textrm{gw+emw}}(\Omega)=
    \frac{1}{2}\,\mathcal{K}(\Omega)
    \left(1-\frac{\sin\Omega\tau}{\Omega\tau}\right)Lh(\Omega),
\end{align*}
which coincides with (\ref{gw_emw_correction}).

\end{document}